\documentclass[aps,prd,reprint,nofootinbib,twoside,notitlepage]{revtex4-1} 
\usepackage{graphicx}
\usepackage{euscript,amssymb}
\usepackage{amsfonts}
\usepackage{amssymb}
\usepackage{amsbsy}
\usepackage{amsmath}
\usepackage{nccmath}   % for de-centering equations environments  i.e.  fleqn environment within which equations envionments are written
\usepackage{color}
\usepackage{bbold}
\usepackage{braket}
\usepackage{dsfont}
\usepackage[T3,T1]{fontenc}
\usepackage[section]{placeins}

\usepackage{tikz}
\usetikzlibrary{snakes}

\usepackage{subcaption} %allows multiple plots in the same figure
\usepackage{mathtools}

\mathtoolsset{showonlyrefs,showmanualtags}

\newcommand{\be}{\begin{equation}}
  \newcommand{\ee}{\end{equation}}
\newcommand{\ben}{\begin{eqnarray*}}
  \newcommand{\een}{\end{eqnarray*}}
\newcommand{\bea}{\begin{eqnarray}}
  \newcommand{\eea}{\end{eqnarray}}
\newcommand{\bdm}{\begin{displaymath}}
  \newcommand{\edm}{\end{displaymath}}
\newcommand{\ba}{\begin{align}}
  \newcommand{\ea}{\end{align}}

\newcommand{\sgn}{\text{sgn}\!}

\DeclareSymbolFont{tipa}{T3}{cmr}{m}{n}
\DeclareMathAccent{\invbreve}{\mathalpha}{tipa}{16}

\begin{document}

\title{Exteriors to bouncing collapse models} 

\author{Tim Schmitz}

\email{tschmitz@thp.uni-koeln.de}

\affiliation{Institut f\"ur Theoretische Physik, Universit\"{a}t zu
K\"{o}ln, Z\"{u}lpicher Stra\ss e 77, 50937 K\"{o}ln, Germany}

\date{\today}

\begin{abstract}

We construct a large class of spacetimes that are smoothly matched to homogeneous, spherically symmetric clouds of matter. The evolution of the clouds is left arbitrary to allow for the incorporation of modifications by quantum effects, which can in particular lead to
bounces. We further discuss two simple yet illustrative examples of these spacetimes, both in general terms and for a specific form of the bounce, with a focus on horizon behavior and relevant timescales. \\
	
\end{abstract}

\maketitle

%%%%%%%%%%%%%%%%%%%%%%%%%%%%%%%%%%%%%%%%%%%%%
%%%%%%%%%%%%%%% INTRODUCTION %%%%%%%%%%%%%%%%
%%%%%%%%%%%%%%%%%%%%%%%%%%%%%%%%%%%%%%%%%%%%%
\section{Introduction}

It is widely believed that the singularities of general relativity will be cured in some form by quantizing the theory. Since such a theory of quantum gravity is not available as of yet, this claim can only be investigated in reduced models constructed in accordance with various approaches to a full theory. Most commonly used are cosmological models and black holes, the latter including both collapse models and eternal black holes.

A result that has emerged in many such investigations in some variation is an avoidance of the classical singularity by a bounce: in cosmological and collapse models, instead of originating or terminating in a singularity, the dynamics transition from collapse to expansion \cite{MeLTB,MeQuantumOS,HajicekKieferNullShells,*KieferNullShellConf,HajicekClassNullShells,*HajicekQuantumNullShells,HajicekNullShells,FrolovNullShell,AlmeidaACS,BergeronBounce,BergeronMixmaster,GozdzBKL,GozdzBKL2,CasadioOS,WilsonQuantumOS,BambiBounce,MalafarinaBounce}. Likewise, eternal black holes have been shown to decay into white holes, see for example Refs.\ \cite{AshtekarBounce,KantowskiSachsBounce}. Although some of these results have recently been called into question \cite{BojowaldCritique,BojowaldBHCritique}, we believe that the pervasiveness of the bounce across various models and approaches to quantum gravity is noteworthy.

Here we are interested in bouncing collapse models. They are typically derived by considering a simple matter distribution, either shells or homogeneous clouds. This then allows for a symmetry reduction of the system, making quantization tractable. The reduction also includes the degrees of freedom of the geometry exterior to the matter configuration, which is then essentially presumed to be completely classical, most often Schwarzschild.

Unfortunately, some of the most interesting questions concerning bouncing collapse involve the exterior; without it we know nothing of the behavior of horizons outside of the matter distribution. How do they transition from trapping to antitrapping, and how long might they be visible to a far-away observer? What does the shadow of such a bouncing compact object look like? These questions have been discussed in the literature on conceptual grounds \cite{AmbrusHajicekLifetime,ChristodoulouLifetime,ChristodoulouLifetime2,BarceloLifetime,BarceloBounce1,BarceloBounce2,BarceloBounce3,LiuBounce,HajicekQuantumNullShells,HaggardRovelliBounce}, but they ultimately require an effective exterior geometry to the bouncing object.

For a more complete review of bouncing collapse and the aforementioned open questions see the review \cite{MalafarinaBounceRev} and references therein. Further we want to note that there have been previous investigations of how quantum effects affect collapse models where the exterior has been explicitly included, see e.g.\ Refs. \cite{WilsonQuantumOS,BojowaldLoopLTB,BojowaldLoopLTB2}.

In this article we want to restrict ourselves to homogeneous dust clouds, described by a Friedman-Lema\^{i}tre-Robertson-Walker (FLRW) geometry, modified by quantum effects to bounce. We present here a large class of exteriors smoothly matched to the dust cloud at its surface. The exact trajectory of the surface is left open, such that these exteriors can be adapted to many bouncing collapse models. Further we discuss two simple yet illustrative examples for such exteriors, both in general and for the specific bouncing trajectory from Refs.\ \cite{MeLTB,MeQuantumOS}.

Previously, similar exteriors have been investigated for bouncing null shells \cite{HaggardRovelliBounce,DeLorenzoFireworks,RovelliFireworks}. Therein regions in the exterior were identified where departures from the classical geometry are unavoidable. These exteriors were further used to investigate the effects of Hawking radiation on the bouncing null shells in Ref.\ \cite{MartinDussaudFireworks}. In Refs.\ \cite{AchourBounce1,AchourBounce2,AchourBounce3} exteriors were discussed in very general terms, allowing modified gravity theories and distributional contributions to the energy-momentum tensor on the matching surface. However, specific examples were restricted to static spacetimes or ones in which no horizons form at all. The same class of static exteriors were investigated also in Ref.\ \cite{MuenchMatching}, where the discussion centered on specific loop quantum gravity models for the bounce. Here we aim to specifically illustrate aspects of dynamic exteriors, with horizons that expand and shrink.

Finally we want to note that these exteriors might share properties with but need not match the various regularized black hole solutions available in the literature, see for example Refs.\ \cite{HaywardRegular,CarballoRubioRegular,AnsoldiRegularRev} . These spacetimes are more or less ad hoc modifications of black hole solutions, smoothing out the geometry near the would-be singularity. This is not necessary for bouncing collapse, since there matter never collapses far enough to reach and uncover these singularities.

We proceed here as follows. In Sec.\ \ref{sec:chapter_4A} we present the construction of our exteriors, and then discuss two specific examples. In Sec. \ref{ch:chapter_5} we investigate these examples further for the specific bouncing trajectory from Refs.\ \cite{MeLTB,MeQuantumOS}, especially with regard to horizons and relevant timescales, before we conclude in Sec.\ \ref{ch:conclusions}.Throughout we use units where $G=c=1$.

%%%%%%%%%%%%%%%%%%%%%%%%%%%%%%%%%%%%%%%%%%%%%%%%%%%%%%%%%%%%%%%%%%%

\section{General construction} \label{sec:chapter_4A}

We assume that the geometry of the interior of the collapsing body can be described by a FLRW line element,
\begin{equation}
	ds^2=-d\tau^2+a^2(\tau)\left(\frac{dr^2}{1-kr^2} +r^2 d\Omega^2 \right) ,
\end{equation} 
where $k\in\{\pm1,0\}$ as usual controls the curvature of spatial slices. The surface of the body can be characterized in the interior by $r=r_S=\text{const.}$ Note that for a closed interior, $k=+1$, one has to restrict $r_S<1$.

It turns out to be convenient to work in adapted coordinates and replace the radial coordinate $r$ by $\rho=r/r_S\leq1$. The line element then takes the form
\begin{equation}
ds^2=-d\tau^2+R_S^2(\tau)\left(\frac{d\rho^2}{1-k_S \rho^2} +\rho^2 d\Omega^2 \right) , \label{eq:line_element_int}
\end{equation} 
where $k_S=k\,r_S^2$ and $R_S(\tau)$ describes the trajectory of the collapsing body's surface at $\rho=1$. At this point we will not restrict $R_S(\tau)$ in any way, and in particular not assume any equations of motion for it. Later we will impose that the collapsing body bounces and expands out again.

To construct exteriors smoothly matched to this interior it turns out to be helpful to first consider a more general metric that contains both the quantum corrected Friedmann model and possible exteriors as special cases, in analogy to the classical Lema\^itre-Tolman-Bondi (LTB) metric. We have previously discussed quantization of the LTB model in Ref.\ \cite{MeLTB}. Hence we choose an ansatz in LTB form,
\begin{equation}
ds^2= -d\tau^2 + \frac{\left( \frac{\partial R}{\partial \rho}\right) ^2}{1 + 2 E(\rho)} d\rho^2 + R^2(\rho,\tau)\,d\Omega^2 . \label{eq:metric_ansatz_ltb}
\end{equation}
We interpret the radial coordinate $\rho$ as is usual for LTB as a label for the spherically symmetric dust shells making up the model, but take it to be rescaled with respect to the surface of the collapsing body in line with Eq.\ \eqref{eq:line_element_int}. It is easy to see that for $\rho<1$ one can reclaim the metric in Eq.\ \eqref{eq:line_element_int} by identifying
\begin{align}
R(\rho,\tau) &= R_S(\tau)\,\rho,\\
E(\rho) &= -\frac{1}{2} k_S \rho^2.
\end{align}

To find an exterior smoothly matched to the interior across $\rho=1$ we hence choose the functions $R(\rho,\tau)$ and $E(\rho)$ for $\rho\geq1$ such that
\begin{align}
R(1,\tau) &= R_S(\tau),  \label{eq:ansatz1} \\
\frac{\partial R}{\partial \tau} &= \mathcal{F}(R_S(\tau),R(\rho,\tau)), \label{eq:ansatz2} \\
E(\rho) &= -\frac{1}{2} k_S \label{eq:ansatz3}, 
\end{align} 
where we have introduced the function $\mathcal{F}$. At this point this function is arbitrary, except at $\rho=1$:
from Eqs.\ \eqref{eq:ansatz1} and \eqref{eq:ansatz2} it follows that there $\mathcal{F}$ is directly determined by the equation of motion,
\begin{align}
\mathcal{F}(R_S(\tau),R(1,\tau))=\left. \frac{\partial R}{\partial \tau} \right|_{\rho=1} =\dot{R}_S, \label{eq:fiducial_identity}
\end{align}
where a dot denotes a derivative with respect to $\tau$, and we assume that the equations of motion for $R_S$ are such that $\dot{R}_S$ can be expressed solely through $R_S$ itself.

The above allows us to make a coordinate transformation introducing the curvature radius $R(\rho,\tau)$ as the radial coordinate,
\begin{equation}
\frac{\partial R}{\partial \rho} d\rho = dR - \frac{\partial R}{\partial \tau} d\tau = dR - \mathcal{F}(R_S(\tau),R) \,d\tau,
\end{equation}
which brings our metric from Eq.\ \eqref{eq:metric_ansatz_ltb} into the form
\begin{multline}
ds^2= -\frac{1 - k_S - \mathcal{F}^2}{1 - k_S} d\tau^2 \\ - \frac{2\mathcal{F}}{1 - k_S} d\tau dR + \frac{dR^2 }{1 - k_S}  + R^2\,d\Omega^2 . \label{eq:line_element_general}
\end{multline}
In App. \ref{ch:chapter_2} we show explicitly that the above and the line element \eqref{eq:line_element_int} are indeed matched smoothly across the dust cloud's surface.

As a consistency check we can see that the above reduces to the Schwarzschild metric for classical Oppenheimer-Snyder collapse, where $\mathcal{F}^2=2M/R - k_S$. We recognize then for $k_S = 0$ the Schwarzschild metric in Painlev\'{e}-Gullstrand coordinates. The same observation also holds for Painlev\'{e}-Gullstrand coordinates generalized to $k_S \neq 0$; compare for $k_S<0$ with Eq. (3.5) in Ref.\ \cite{MartelCoordinates} (apart from a constant rescaling of $\tau$), where we can identify $p=1/(1-k_S)$, and for $k_S>0$ with Eq.\ (10) in Ref.\ \cite{GautreauCoordinates}, with the identification $R_i=2M/k_S$. Note that the latter coordinates are only valid for $R<R_i<2M$.

There is of course a great amount of freedom in choosing the function $\mathcal{F}$, since the matching only determines it on the dust cloud's surface. For the remainder of this section we will explore some specific choices for this function, and see what kind of exterior they lead to. Furthermore we want to note that we did not choose the most general ansatz for exteriors. One could generalize $E$ to vary with $\rho$ and only fulfill Eq.\ \eqref{eq:ansatz3} at $\rho=1$. We have not done so here because such exteriors turn out to be somewhat further removed from Schwarzschild when inserting specific equations of motion.

%%%%%%%%%%%%%%%%%%%%%%%%%%%%%%%%%%%%%%%%%%%%%%%%%%%%%%%%%%%%%%%%%%%%%%%%%

\subsection{Static exteriors} \label{ch:chapter_3}

Let us now be more specific and assume that the quantum corrected equation of motion can be given in the form
\begin{align}
	\dot{R}_S^2&=F(R_S),  \label{eq:form_of_eom1}\\
	\ddot{R}_S&=\frac{1}{2} F'(R_S). \label{eq:form_of_eom2}
\end{align}
Since we want the resulting trajectory to bounce at some minimal radius $R_0$, where $\dot{R}_S = 0$ and $\ddot{R}_S>0$, we can impose $F(R_0)=0$ and $F'(R_0)>0$. Further assuming that this trajectory approaches the classical one far away from the singularity we can impose that $F(R_S)\sim2M/R_S - k_S$ for large $R_S$, where $M$ is the total mass of the dust cloud.

First we want to consider a static exterior analogous to a Schwarzschild black hole, hence we choose
\begin{equation}
	\mathcal{F}(R_S(\tau),R)=\mathcal{F}(R).
\end{equation}
$\mathcal{F}$ is then immediately determined by the equation of motion,
\begin{equation}
	\mathcal{F}(R)=\eta\sqrt{F(R)},
\end{equation}
where $\eta=\sgn \,\dot{R}_S$. Despite the appearance of $\eta$ the resulting exterior geometry is still static. The easiest way to see that is to bring the line element into the form
\begin{equation}
	ds^2=-f(R) \,dT^2 + \frac{dR^2}{f(R)} + R^2 d\Omega^2 \label{eq:static}
\end{equation}
with the help of the coordinate transformation
\begin{equation}
	\tau=\sqrt{1-k_S}\,T-\eta\int dR~\frac{\sqrt{F(R)}}{f(R)},
\end{equation}
where
\begin{equation}
	f(R) = 1 - F(R) - k_S .
\end{equation}

We can immediately say that the resulting exterior spacetime approaches Schwarzschild for large curvature radii. How far away from the collapsing body the quantum corrections are still noticeable depends on the specific quantum corrected equation of motion.

Of particular interest is the emergence of horizons in this exterior. To this end we follow Ref.\ \cite{FaraoniHorizons} and identify apparent horizons by $f(R)=0$. They separate untrapped regions, where $f(R)>0$, from (anti-)trapped regions, where $f(R)<0$. In the following we will assume that all real roots of $f(R)$ are simple. From the behavior of $F(R)$ discussed above we can say that at the radius of the bounce we have
\begin{equation}
	f(R_0) = 1 - k_S > 0,
\end{equation}
and hence the bounce takes place in an untrapped region. The region $R<R_0$ is of no further importance here, since it is always covered by the collapsing dust cloud. Due to the asymptotic behavior of $F(R)$ the exterior is asymptotically flat and hence untrapped at large $R$ as well. Depending on the number of roots of $f(R)$ these two untrapped regions might be separated from each other by alternating (anti-)trapped and untrapped regions.

These results match the more general discussion in Ref.\ \cite{AchourBounce1,AchourBounce2,AchourBounce3}. There it was found for a much wider range of models that the bounce always takes place in an untrapped region of the exterior, making it necessary for a potential outer horizon always to be paired up with an inner horizon.

Since we want to compare the comoving with the exterior observer, we want to find the trajectory of the dust cloud's surface in Killing time $T$ given by:
\begin{equation}
	\left(\frac{d\overline{R}}{dT} \right)^2 = \frac{\dot{R}^2_S}{\dot{\overline{T}}^2} = F(\overline{R})^2 \frac{f(\overline{R})^2}{1-k_S},
\end{equation}
where an overline denotes a quantity on the surface of the dust cloud, such that for a function $q(R,\tau)$ we have $\overline{q}=q(R_S(\tau),\tau)$.
Expanding this equation of motion near the horizons, should there be any, we can say $f(\overline{R})=(\overline{R}-R_\text{h})\tilde{f}(R_\text{h})$ with $\tilde{f}(R_\text{h})\neq0$, since the roots of $f(R)$ are simple. Keeping in mind that $f(R)$ and $F(R)$ differ by an additive constant and can thus never share roots we can then say that near the horizons
\begin{equation}
	\left(\frac{d\overline{R}}{dT} \right)^2 \propto (\overline{R}-R_\text{h})^2 .
\end{equation}
We can then read off that $\overline{R}(T)$ only approaches the horizons asymptotically and never crosses them in finite Killing time, in analogy with a Schwarzschild exterior. 

The fact that the exterior observer then never observes the bounce itself when horizons are present, but the dust cloud in fact expands out again hints toward the necessity of the existence of other asymptotically flat regions in the maximal extension of this static exterior; the dust cloud then reemerges in a "parallel universe".

This we will discuss in more detail in Sec.\ \ref{ch:chapter_5} at the example of a particular quantum corrected equation of motion. Another example for this behavior can also be found in Ref.\ \cite{MuenchMatching}, illustrated with the Penrose diagram in Fig.\ (14).

While this construction is in principle a consistent description of bouncing collapse, we find the necessity of a parallel universe where the expansion takes place undesirable. It seems thus unavoidable to consider dynamic exteriors to have both a bounce and horizons. This assertion that there should be a unique asymptotic region has also been expressed in Ref. \cite{HajicekClassNullShells}, although from a technical rather than conceptual standpoint.

Before moving on to dynamic exteriors we want to emphasize that we do not exclude the possibility of consistent static exteriors completely. By relaxing the assumption that the roots of $f(R)$ be simple it might be possible for the exterior to have apparent horizons that can be crossed in finite Killing time. Should one insist on a static exterior, this can be understood as a restriction on possible quantum corrected equations of motion.

%%%%%%%%%%%%%%%%%%%%%%%%%%%%%%%%%%%%%%%%%%%%%%%%%%%%%%%%%%%%%%%%%%%%%%%%%

\subsection{Time-dependent-mass exteriors} \label{sec:chapter_4B}
To overcome the problems of static exteriors we want to construct a possible dynamic exterior. Here we will consider a very simple and hopefully instructive case. We do not claim that this specific exterior gives a realistic model of a bouncing black hole. 
In fact as we will see shortly, since its ADM mass changes in time and even vanishes at the moment of the bounce it will most likely exhibit some of the undesirable properties described in Refs.\ \cite{FaraoniUnphysical1,FaraoniUnphysical2}.
We still believe that a closer investigation of this simple example will be helpful for later, more systematic searches for a consistent bouncing black hole model.

The smaller class of solutions we consider here are restricted to $k_S=0$ and fulfill
\begin{equation}
	\mathcal{F}(R_S(\tau),R) = \eta \sqrt{\frac{2 M(R_S(\tau))}{R}},
\end{equation}
where $\eta = \sgn\,\dot R_S$. These solutions describe a generalization of Schwarzschild in Painlev\'{e}-Gullstrand form, where the mass varies with comoving time. It follows from Eq.\ \eqref{eq:fiducial_identity} that this mass is determined by the quantum corrected equation of motion as
\begin{equation}
	M(R_S) = \frac{1}{2} R_S \dot{R}_S^2. \label{eq:def_mass_t}
\end{equation}
In Painlev\'{e}-Gullstrand form the metric is then
\begin{multline}
	ds^2= -\left( 1 - \frac{2 M(R_S)}{R}\right)  d\tau^2 \\ - 2\eta\sqrt{\frac{2 M(R_S)}{R}} d\tau dR + dR^2  + R^2\,d\Omega^2 . \label{eq:line_element_mass_t} 
\end{multline}

We consider here only the flat case for simplicity. A generalization especially to the closed case $k_S>0$ requires some care, since it needs to involve an extension of the $(\tau,R)$ coordinates past the aforementioned restriction $R<2M(R_S)$.

First we want to discuss apparent horizons. Following Ref.\ \cite{FaraoniHorizons}, the expansions of outgoing null geodesics $\Theta_+$ and ingoing null geodesics $\Theta_-$ are given by
\begin{equation}
	\Theta_\pm = \pm\frac{2}{R}\left(1 \pm \eta \sqrt{\frac{2 M(R_S)}{R}} \right).
\end{equation}
There is thus only one apparent horizon, determined by $\Theta_+=0$ or $\Theta_-=0$, at $R=2M(R_S)$, in contrast to the static exteriors discussed in the last section. It separates the untrapped asymptotic region $R>2M(R_S)$, where $\Theta_+>0$ and $\Theta_-<0$, from the region $R<2M(R_S)$ that is either trapped or antitrapped depending on $\eta$: before the bounce, $\eta=-1$, this region is trapped since $\Theta_\pm<0$, and after the bounce, $\eta=+1$, it is antitrapped since $\Theta_\pm>0$. The transition from trapped to antitrapped is facilitated by the horizon withdrawing into the origin $R=0$ at the instant of the bounce,
\begin{equation}
	2M(R_S)=R_S\dot{R}_S^2\overset{\dot{R}_S=0}{=} 0.
\end{equation}

Whether the horizon is outside of the collapsing body at any given time can be determined by the sign of the function $R_S - R_S\dot{R}_S^2$. The attentive reader might have spotted that this is the same function we used to determine whether a region in the static exterior is (anti-) trapped or untrapped in Sec.\ \ref{ch:chapter_3}. We can thus apply those results here, making the same assumptions about $\dot{R}_S$ given by Eqs.\ \eqref{eq:form_of_eom1} and \eqref{eq:form_of_eom2}: For early and late times, where $R_S(\tau)$ is large, the body's surface is outside of the horizon. Approaching the bounce, surface and horizon cross through each other an even number of times such that at the bounce the horizon is inside of the collapsing body. Note that a similar picture has also emerged in Refs. \cite{BambiBounce,MalafarinaBounce}, although there the exterior metric has not been given explicitly.

Lastly we want to discuss two more properties of this exterior: its matter content and the Kretschmann scalar to check for curvature singularities.
We start with the latter. It is given by
\begin{align}
	\mathcal{K}&=\mathcal{R}^{\mu\nu\rho\lambda}\mathcal{R}_{\mu\nu\rho\lambda}\\&=\frac{48M^2}{R^6} -  \frac{24 M}{\sqrt{R^9 R_S}}\,\frac{\partial M}{\partial R_S} + \frac{9}{R^3 R_S}\,\left( \frac{\partial M}{\partial R_S}\right)^2 ,
\end{align}
where $\mathcal{R}_{\mu\nu\rho\lambda}$ is the Riemann tensor. As is apparent, there is a singularity at $R=0$. This is of no further importance, since when matched with the bouncing interior this singularity never appears in the full spacetime. Since for a bouncing collapse $R_S$ never vanishes, the Kretschmann scalar does not diverge anywhere else as long as the derivative of $M$ is well behaved. Through Eq.\ \eqref{eq:def_mass_t} this derivative can be found as
\begin{equation}
	\frac{\partial M}{\partial R_S} = \frac{1}{2}\left(\dot{R}_S^2 + 2 R_S \ddot{R}_S \right),  \label{eq:mass_tau}
\end{equation}
hence it is not too much of a restriction on the quantum corrected equation of motion that this should stay finite.

Computing the Einstein tensor and imposing the Einstein field equations we further find that the energy momentum tensor generating this exterior can be expressed as 
\begin{equation}
	T_{\mu\nu} = p_t ( u_\mu u_\nu + g_{\mu\nu}) + (p_r - p_t) n_\mu n_\nu ,
\end{equation}
where $u_\mu dx^\mu=d\tau$ is the unit co-vector in direction of the comoving time and
\begin{equation}
	n_\mu dx^\mu = dR-\eta\sqrt{\frac{2M}{R}}d\tau
\end{equation}
is the unit co-vector normal to the collapsing body's surface. The quantities $p_r$ and $p_t$ can thus be interpreted as pressures radial and tangential to the surface, respectively, and are given by
\begin{align}
	8\pi p_r = -\frac{2}{R^2}\,\sqrt{\frac{R}{R_S}}\, \frac{\partial M}{\partial R_S},\\
	8\pi p_t = -\frac{1}{2R^2}\,\sqrt{\frac{R}{R_S}}\,\frac{\partial M}{\partial R_S}.
\end{align}
The matter content can thus be regarded as an ideal fluid with vanishing energy density and anisotropic pressure. As one can see from Eq.\ \eqref{eq:mass_tau}, at the bounce, where $\dot{R}_S=0$ and $\ddot{R}_S>0$, the pressures are negative. This matches our results concerning the effective matter of quantum corrected LTB collapse in \cite{MeLTB}, see also Ref.\ \cite{MalafarinaBounce}. As already discussed there, the violation of various energy conditions is an advantage rather than a flaw of the model, since it allows us to evade the Penrose-Hawking singularity theorems.

Just as the Kretschmann scalar, the energy momentum tensor is well behaved. In conclusion we can say that this dynamic exterior seems to be free of possible pathologies in these regards and also evades the unfavorable causal structure of the static exterior. Below we will explore it in more detail by specifying a quantum corrected equation of motion.

There we will also discuss how the bounce looks from the perspective of an exterior observer. Of special importance is the time that the horizons are visible for. This is not straightforward in our non-static exterior, since with staticity we have lost an important criterion to single out the exterior observer. To circumvent this problem, we will make use of a more operational standpoint to determine the lifetime. For details, see Sec.\ \ref{sec:chapter_5B}.

%%%%%%%%%%%%%%%%%%%%%%%%%%%%%%%%%%%%%%%%%%%%%%%%%%%%%%%%%%%%%%%%%%%%%%%%%

\section{a specific equation of motion} \label{ch:chapter_5}
For the remainder of this article we want to focus on one specific quantum corrected equation of motion for $k_S=0$,
\begin{equation}
	\dot{R}_S^2= \frac{2M_0}{R_S}\left(1-\frac{R_0^3}{R_S^3} \right), \label{eq:specific_eom}
\end{equation}
with the solutions
\begin{equation}
	R_S(\tau)=\left[R_0^3+\frac{9M_0}{2}(\tau-\tau_0)^2 \right]^\frac{1}{3}, \label{eq:specific_trajectory}
\end{equation}
where $M_0$ is the initial total mass of the collapsing body and $R_0$ the minimal radius of the bounce reached at $\tau=\tau_0$. We have found and discussed this equation and its solutions in Refs.\ \cite{MeOS,MeQuantumOS}, where we constructed a quantum Oppenheimer-Snyder model, and also in Ref.\ \cite{MeLTB} where it emerged for a quantum Lema\^{i}tre-Tolman-Bondi model. There we have also seen that $R_0^3=\hbar^2\delta/M_0$, where $\delta$ is a parameter determined by quantization ambiguities.

\subsection{Static exterior} \label{sec:chapter_5A}
Following the general procedure laid out in Sec.\ \ref{ch:chapter_3}, the static exterior corresponding to Eq.\ \eqref{eq:specific_eom} is given by the line element \eqref{eq:static} with
\begin{align}
	f(R) = 1-\frac{2M_0}{R}\left(1-\frac{R_0^3}{R^3} \right). \label{eq:static_f}
\end{align}
To find the horizons of this exterior, and with this its causal structure, we need to find the roots of $f(R)$.

To this end we note that for both $R\to0$ and $R\to\infty$ the function $f(R)$ is positive, and that it has a local minimum at $R=2^\frac{2}{3}R_0$ where it takes the value
\begin{equation}
	f(2^\frac{2}{3}R_0) = 1-\frac{3 M_0}{2^\frac{5}{3}R_0}.
\end{equation}
$f(R)$ has real roots only when this value is non-positive. Thus we see that for $2^\frac{5}{3}R_0>3 M_0$ there are no roots and hence the exterior has no horizons. 

\begin{figure}
	\centering
	\includegraphics[width=0.45\textwidth]{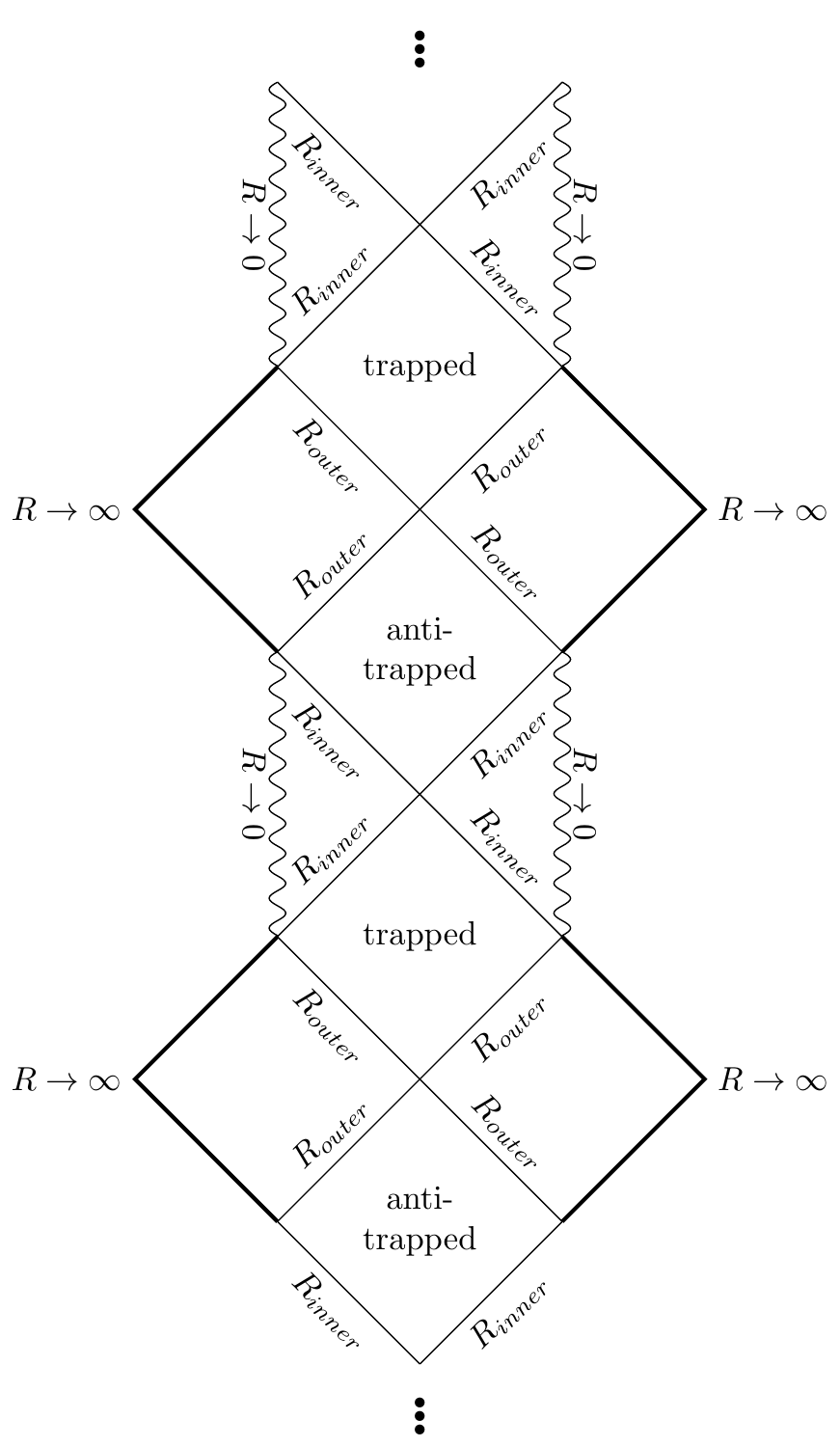}
	\caption{Penrose diagram for the static exterior given by Eq.\ \eqref{eq:static_f} with $2^\frac{5}{3}R_0<3 M_0$. Thick lines denote null infinities, thin lines horizons and wavy lines singularities.}
	\label{fig:penrose}
\end{figure}

More interesting is the case $2^\frac{5}{3}R_0<3 M_0$, for which $f(R)$ has two roots and thus two horizons, inner and outer, emerge. The inner horizon's position we can estimate as $R_\text{inner}<2^\frac{2}{3}R_0<\frac{3}{2}M_0$, and the outer horizon's position, noting $f(2M_0)>0$, as $2^\frac{2}{3}R_0<R_\text{outer}<2M_0$. This configuration of the horizons is reminiscent of that of a Reissner-Nordstr\"om black hole, and so is its causal structure. We can illustrate this with the Penrose diagram for this exterior, see Fig.\ \ref{fig:penrose}. Details of its construction following Ref.\ \cite{SchindlerDiagram} can be found in App.\ \ref{app:A}.

For $2^\frac{5}{3}R_0=3 M_0$, when $f(R)$ has a single root, the exterior assumes an extremal configuration in further analogy with the Reissner-Nordstr\"om black hole. Since this root is not simple anymore, discussion of the causal structure is more complicated and will not be undertaken here.

In Ref.\ \cite{MuenchMatching}, Fig.\ (14), a comparable Penrose diagram was found, although there the black hole and white hole blocks in the diagram are condensed into a single block with a transition surface in between, and no singularities are present.

\subsection{Time-dependent-mass exterior} \label{sec:chapter_5B}

Following Sec.\ \ref{sec:chapter_4B}, we can construct a dynamic exterior with line element
\begin{multline}
	ds^2= -\left( 1 - \frac{2 M(R_S)}{R}\right)  d\tau^2 \\ - 2\eta \sqrt{\frac{2 M(R_S)}{R}} d\tau dR + dR^2  + R^2\,d\Omega^2 ,
\end{multline}
where the time-dependent mass is
\begin{equation}
	M(R_S(\tau)) = M_0\left(1-\frac{R_0^3}{R_S^3(\tau)} \right). \label{eq:Mass}
\end{equation}
For $\tau\to\pm\infty$, where $R_S\to\infty$, we have $M(R_S)\to M_0$. For early and late times, away from the bounce, this exterior is thus approximately Schwarzschild with mass $M_0$. When the collapsing body's surface approaches the minimal radius, $M(R_S)$ decreases until it vanishes at the time of the bounce and increases again during expansion.

\subsubsection{Horizons}

With the mass also the horizon at $R=2M(R_S)$ first contracts and then expands again. To determine whether the horizon is outside of the body's surface at any given time we can, as discussed in Sec.\ \ref{sec:chapter_4B}, largely adapt our analysis of the horizons in the last section: For $2^\frac{5}{3}R_0>3 M_0$ the horizon never emerges from the collapsing matter. For $2^\frac{5}{3}R_0<3 M_0$ it does emerge during the collapse when $R_S(\tau)=R_\text{outer}$, where $2^\frac{2}{3}R_0<R_\text{outer}<2M_0$,  and disappears again at $R_S(\tau)=R_\text{inner}<2^\frac{2}{3}R_0$. During the expansion this is repeated in reverse.

\begin{figure}
	\centering
	\includegraphics[width=0.45\textwidth]{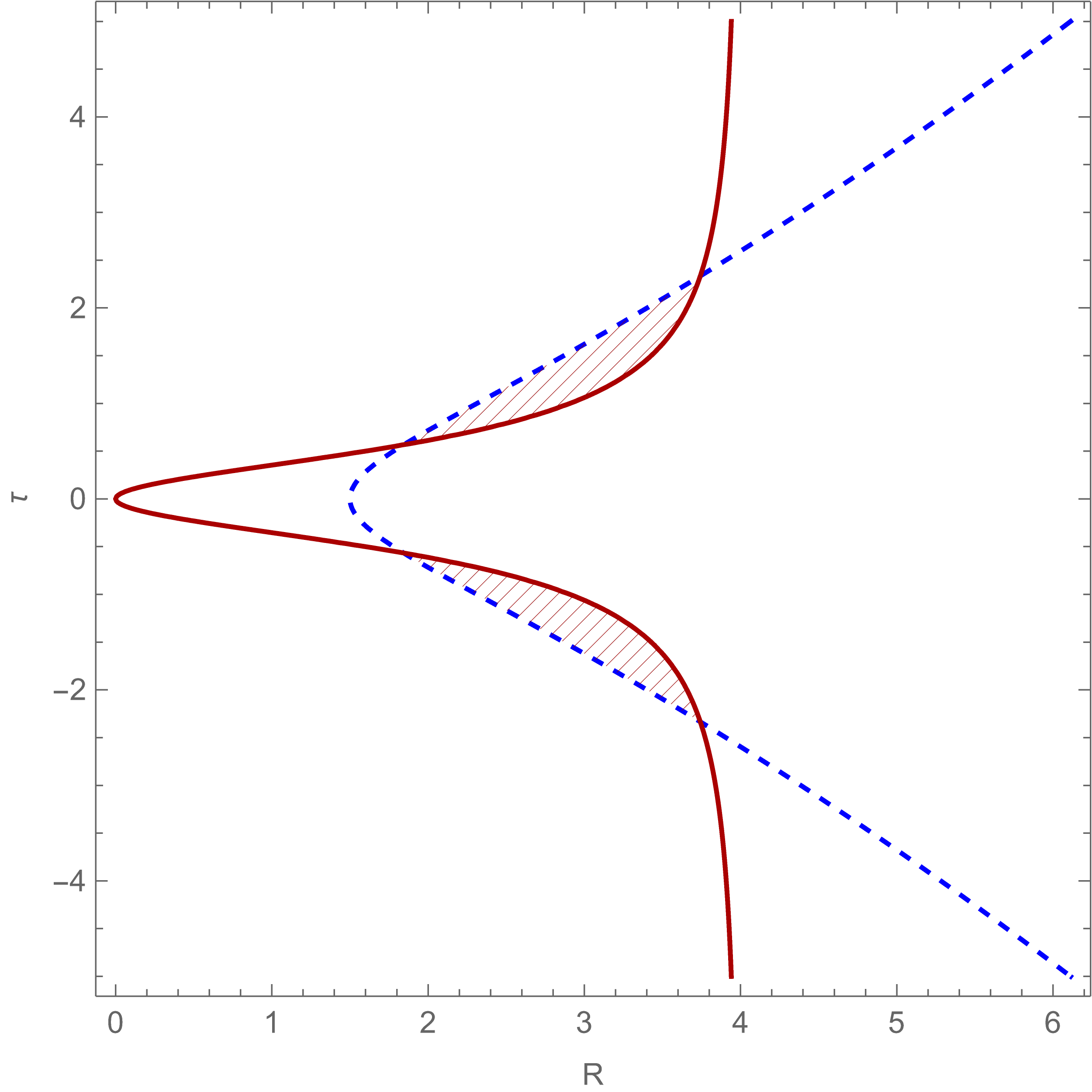}
	\caption{The quantum corrected trajectory given by Eq.\ \eqref{eq:specific_trajectory} (dashed blue line) as compared to the horizon $R=2M(R_S(\tau))$ as given by Eq.\ \eqref{eq:Mass} (full red line) for $R_0=1.5$ and $M_0=2$ in Planck units. Hatching denotes the (anti-)trapped regions in the exterior.}
	\label{fig:horizons}
\end{figure}

We illustrate this process as seen by the comoving observer schematically in Fig.\ \ref{fig:horizons}, using $\tau$ as the time coordinate. Note that for astrophysical scales the disappearance and reemergence of the horizon, its transition from black hole to white hole horizon, could happen much more rapidly than the figure suggests. A convenient notion for the timescale of this process is the duration that $R_S(\tau)<2^\frac{2}{3}R_0$, since from previous considerations we know that at $R_S(\tau)=2^\frac{2}{3}R_0$  the horizon is always in the exterior. A short calculation gives for this timescale
\begin{equation}
	\Delta\tau_{\text{trans}} = \sqrt{\frac{8R_0^3}{3M_0}}.
\end{equation}
Assuming a solar mass collapsing body we find that for $R_0$ at the Planck scale, $\Delta\tau_{\text{trans}}\sim10^{-19}t_p\sim10^{-44}s$. The transition of the horizon would then take place on a sub-Planckian timescale. For $\Delta\tau_{\text{trans}}$ to be higher one also has to choose a larger $R_0$, but there is an upper bound: since $2^\frac{5}{3}R_0<3 M_0$, we have $\Delta\tau_{\text{hor}} < \frac{3}{2}M_0$. For $M_0$ again being the solar mass this upper bound is of the order of microseconds.

\begin{figure}
	\centering
	\begin{subfigure}{0.45\textwidth}
		\centering
		\includegraphics[width=\textwidth]{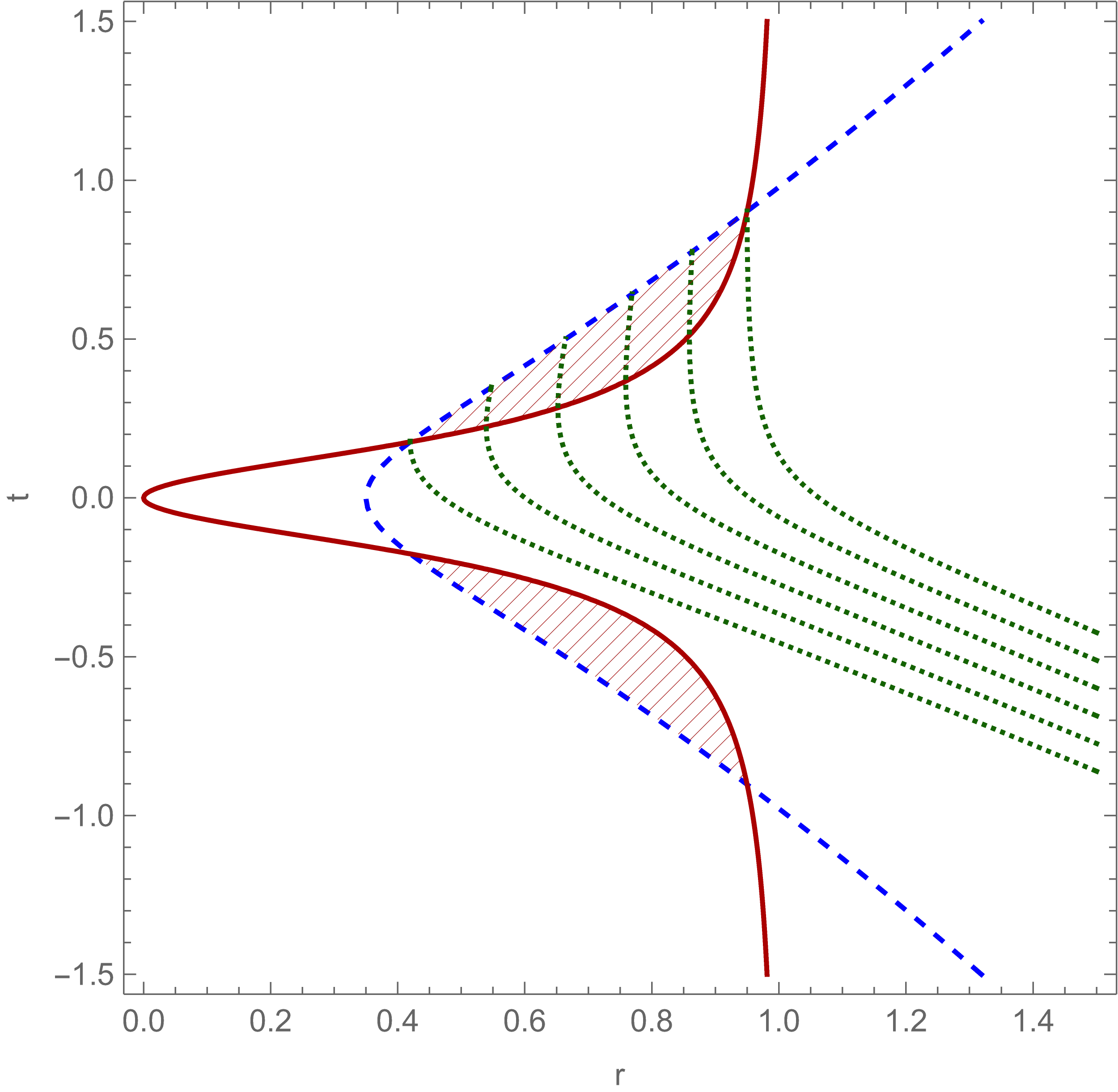}
		\caption{$\frac{R_0}{2M_0}=0.35$}
	\end{subfigure}
	\begin{subfigure}{0.45\textwidth}
		\centering
		\includegraphics[width=\textwidth]{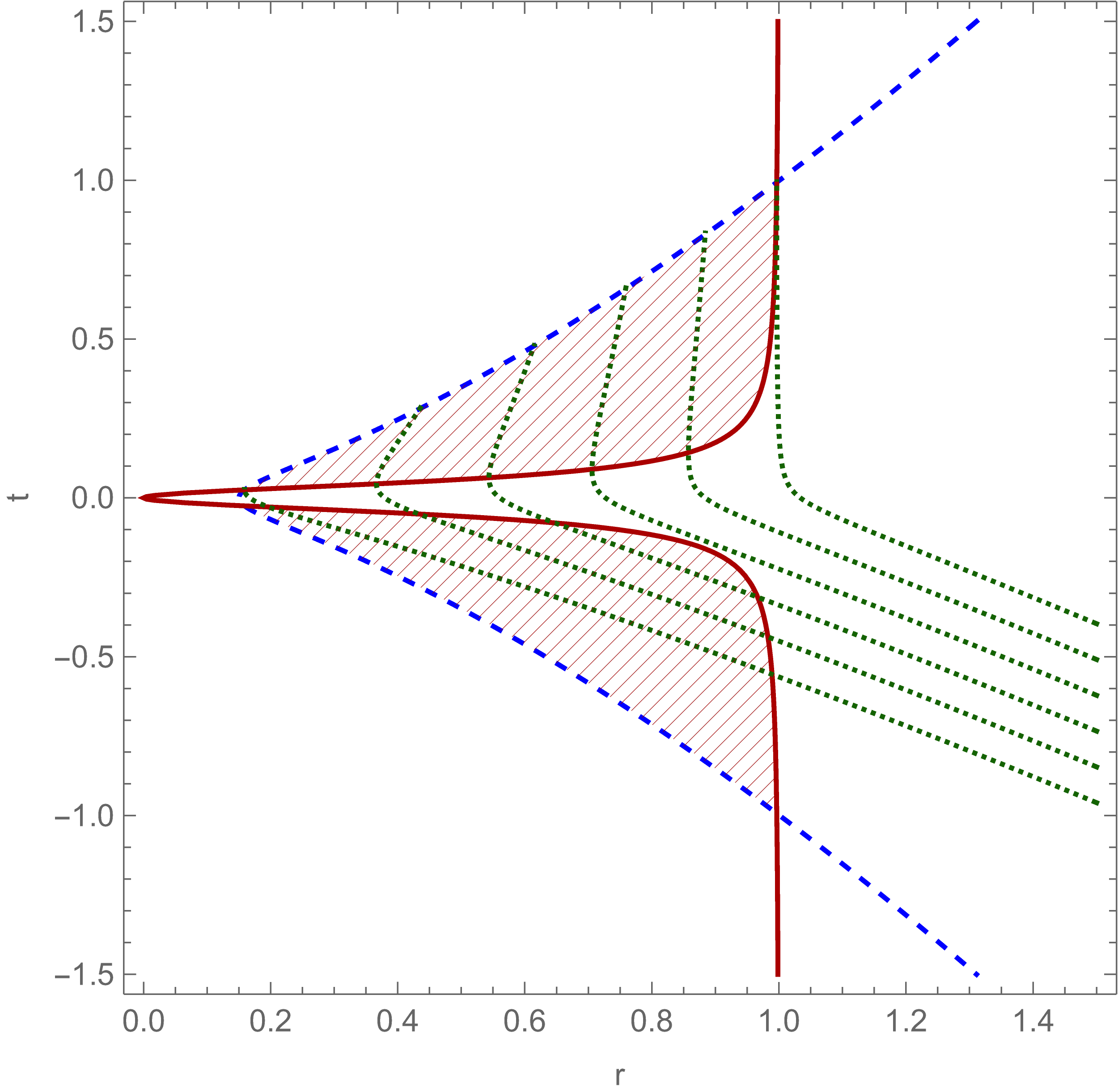}
		\caption{$\frac{R_0}{2M_0}=0.15$}
	\end{subfigure}
	\caption{The rescaled quantum corrected trajectory $R_S/2M_0$, see Eq.\ \eqref{eq:specific_trajectory}, (dashed blue line) as compared to the horizon (full red line) and a family of ingoing light rays (dotted green lines) for different values of $R_0/2M_0$. Hatching denotes the (anti-)trapped regions in the exterior.}
	\label{fig:instability}
\end{figure}

We want to note here that light rays can still escape the trapping region or penetrate the antitrapping region due to the movement of the horizon. We illustrate this in Fig.\ \ref{fig:instability}, where we plotted numerically computed ingoing null geodesics in our exterior; the lightrays momentarily stop at the antitrapping horizon, but the outwardly expanding horizon swallows them up regardless. Light rays emitted from the collapsing body's surface inside of the trapping region can escape to infinity in the same way. Note that in Fig.\ \ref{fig:instability} we use rescaled quantities $t$ and $r$ defined by $4M_0t=3\tau$ and $R=2M_0 r$. More on null geodesics later.

It would certainly be interesting to investigate how this behavior influences possible observational signatures of this bouncing black hole: do closed geodesics behave similarly, and if yes how does this imprint on the black hole shadow? Do signals escaping from the trapped region show special characteristics that could be identified in astrophysical data? We leave a closer discussion of this for future work. Here we only want to note that any observational signatures should be connected to $\Delta\tau_{\text{trans}}$, since the horizons still trap lightrays when they are close to stationary.

\subsubsection{Black hole lifetime}

What we want to investigate in the following is the black hole lifetime. Note that this lifetime is different from the $\Delta\tau_{\text{trans}}$ discussed above; $\Delta\tau_{\text{trans}}$ tells us how rapid the transition from black hole to white hole is. The lifetime we discuss now instead determines how long the (anti-)trapped regions are present in the exterior at all, from the viewpoint of the exterior observer. What characterizes this observer is that they remain at a fixed curvature radius, and that their proper time coincides with our comoving time $\tau$ when this radius is taken to infinity.

To compute the lifetime of the horizon for this observer we hence make the following construction. We identify the first and last moment where the horizon exists by $R_S(\pm\tau_\text{outer})=R_\text{outer}$, where $\tau_\text{outer}>0$. Then we trace ingoing light rays from these events backwards in time, and determine at which times $\tau_1$ and $\tau_2$, $\tau_1<\tau_2$, those two light rays originated from a fixed $R_\text{obs}$. We then find the lifetime as 
\begin{equation}
	\Delta\tau_\text{ext}=\lim_{R_\text{obs}\to\infty}\tau_2-\tau_1. \label{eq:lightrays}
\end{equation} 
This lifetime hence roughly speaking characterizes for how long light rays are absorbed by the collapsing object's horizon.

When one parametrizes null geodesics in this exterior by comoving time $\tau$, they can be described by $R_\pm(\tau)$ fulfilling
\begin{align}
	\dot{R}_\pm &= \frac{R_\pm}{2}\Theta_\pm =\eta \sqrt{\frac{2M(R_S)}{R_\pm}}\pm 1\\
	&= \frac{3M_0\tau}{\sqrt{R_\pm}}\frac{1}{\sqrt{R_0^3+\frac{9M_0}{2}\tau^2}} \pm 1,
\end{align}
where the upper sign denotes outgoing light rays and the lower sign ingoing ones. For our purposes here the latter suffices. As mentioned above, it turns out to be convenient to rescale the quantities involved as $R_-=2M_0 r$ and $4M_0t=3\tau$, which gives
\begin{equation}
	\frac{3}{2}\frac{d r}{d t} = \frac{t}{\sqrt{r}}\frac{1}{\sqrt{\big( \frac{R_0}{2M_0}\big)^3 + t^2 }} - 1.
\end{equation}
We see that in this form of the equation, the two free parameters from Eq.\ \eqref{eq:specific_eom} $R_0$ and $M_0$ only enter as their quotient.

To follow our construction as outlined above, we now have to find two solutions $r_1(t)$ and $r_2(t)$ to this equation, respectively with the initial (or rather final) conditions $r_1(-t_\text{outer})=r_\text{outer}$ and $r_2(t_\text{outer})=r_\text{outer}$. Then we find $t_1$ and $t_2$ from which the lifetime is inferred from $r_1(t_1)=r_2(t_2)=r_\text{obs}$ for $r_\text{obs}\to\infty$. Unfortunately this cannot be done analytically. 

We can, however, estimate the result when $R_0/2M_0$ is small. As noted before, in our quantum Oppenheimer-Snyder model the minimal radius is given by $R_0^3 = \hbar^2\delta/M_0$, where $\delta$ is determined by quantization ambiguities. Hence this estimate can be understood either as approaching the classical limit $\hbar\to0$, or the limit of large masses $M_0\to\infty$. 

Eq.\ \eqref{eq:lightrays} can then be approximated to first order as
\begin{equation}
	\frac{3}{2}\frac{d r}{d t} \approx \frac{\sgn~t}{\sqrt{r}} - 1. \label{eq:lightrays_est}
\end{equation}
Further we find $R_\text{outer}=2M_0$ and $\tau_\text{outer}=4M_0/3$, giving us $r_\text{outer}=1$ and $t_\text{outer}=1$. In this limit one can thus imagine our full bouncing collapse model, interior with dynamic exterior, as classical Oppenheimer-Snyder collapse reaching the singularity at $\tau=0$, glued to a time reversed copy of itself across $\tau=0$. Our full model with $R_0\neq0$ can then be understood as a smoothing out of this very primitive, quasi-classical model for bouncing collapse.

Let us now consider our two lightrays: the first reaches $r=1$ at $t=-1$ unimpeded. The second lightray is more interesting. Since for $t>0$ the horizon is antitrapping, lightrays can only approach it asymptotically. The only way an ingoing lightray can then reach $r=1$ at $t=1$ is for it to get caught on the horizon just as it transitions from trapping to antitrapping. The lightray we are looking for thus reaches $r=1$ already at $t=0$. When we restrict to $t<0$, Eq.\ \eqref{eq:lightrays_est} does not directly depend on $t$. Hence both lightrays follow the same trajectory, just shifted in $t$, when approaching $r=1$ from infinity. From this follows directly that this time shift between the two trajectories at equal $R$ remains constant. We can hence conclude that in this limit the lifetime is given by $\Delta t_\text{ext} \approx 1$, or
\begin{equation}
	\Delta\tau_\text{ext} \approx \frac{4 M_0}{3}. \label{eq:lifetime_result}
\end{equation} 

The discussion above additionally implies that $\Delta\tau_\text{ext}$ defined in this way is not very sensitive to the antitrapping phase of the horizon: if the expansion of the collapsing object is delayed and the antitrapping horizon is present in the exterior for a longer time, the second lightray is simply stuck on the horizon for longer. The lifetime is then not affected, at least in the limit we are currently considering.

One can analogously define a $\Delta\tau_\text{ext}$ that is more sensitive to the antitrapping horizon by using outgoing instead of ingoing lightrays. Of course this does not make a difference for our purposes here, since our exterior is symmetric with respect to time reversal.

\begin{figure}
	\centering
	\vspace{2em}
	\includegraphics[width=0.45\textwidth]{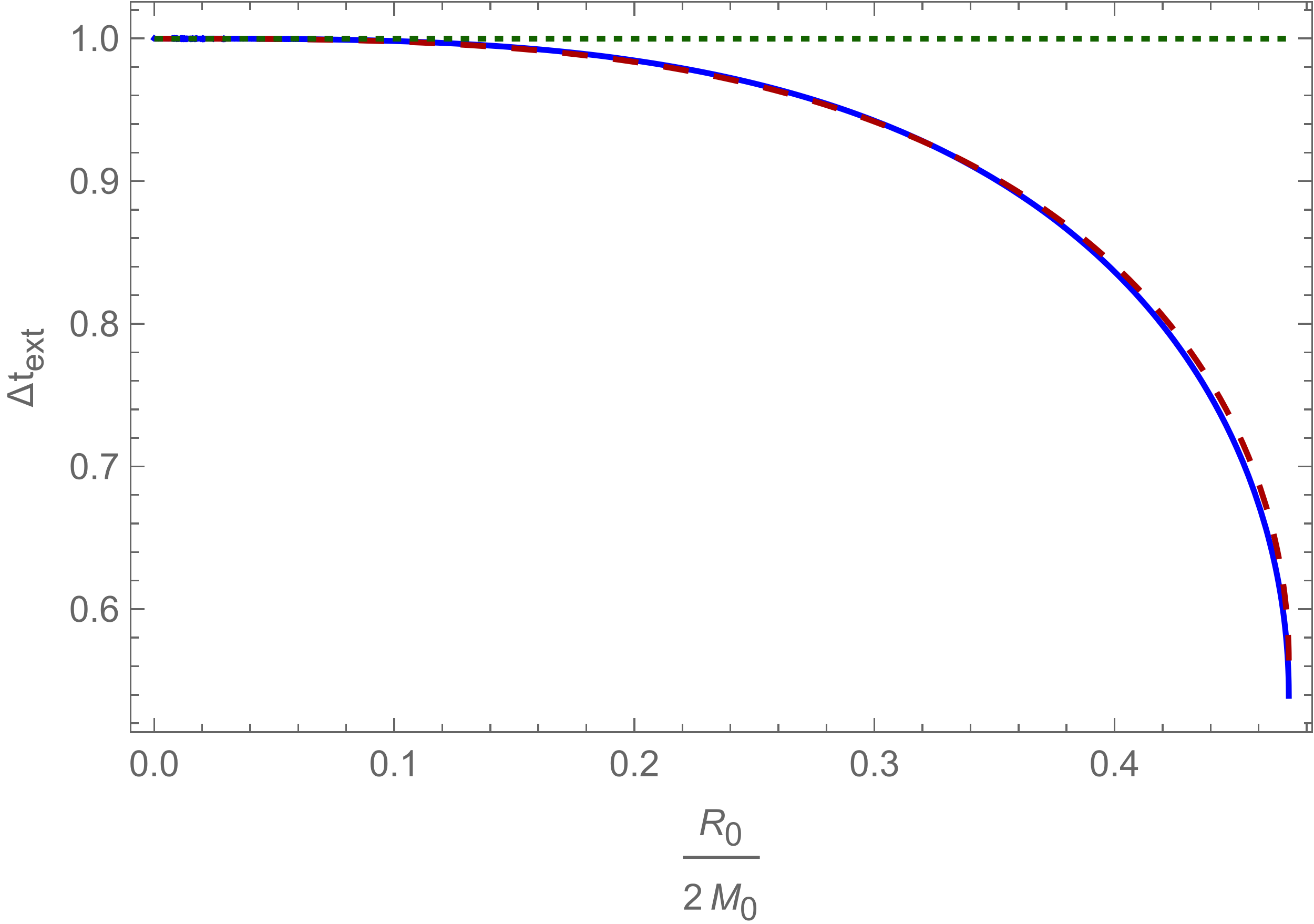}
	\caption{The rescaled lifetime $\Delta t_\text{ext}$ (full blue line) as a function of $R_0/2M_0$ for $r_\text{obs}=10^3$, compared to the approximated result $\Delta t_\text{ext}=1$ (dotted green line) and $t_\text{outer}$ (dashed red line).}
	\label{fig:lifetime}
\end{figure}

Investigating the lifetime numerically, starting from the full null geodesic equation \eqref{eq:lightrays}, confirms that our result \eqref{eq:lifetime_result} is valid for low $R_0/2M_0$, see Fig.\ \ref{fig:lifetime}. When $R_0/2M_0$ approaches its maximum value $3\cdot2^{-8/3}\approx 0.47$, the lifetime even decreases further.

To understand how this happens it is useful to distinguish between two different contributions to the lifetime, both depending on $R_0/2M_0$: firstly, non-zero values of this parameter cause lightrays originating at the trapping horizon to escape earlier, since the horizon moves inwards. This increases the lifetime. Secondly, increasing $R_0/2M_0$ decreases $t_\text{outer}$, the time where the horizon disappears back into the dust cloud. This decreases the lifetime, since the two lightrays are emitted in shorter succession. Comparing $t_\text{outer}$ with $\Delta t_\text{ext}$ clearly shows that the second contribution is much more relevant than the first one, overall leading to a drastic decrease in lifetime for higher $R_0/2M_0$.

\begin{figure}
	\centering
	\begin{subfigure}{0.45\textwidth}
		\centering
		\includegraphics[width=\textwidth]{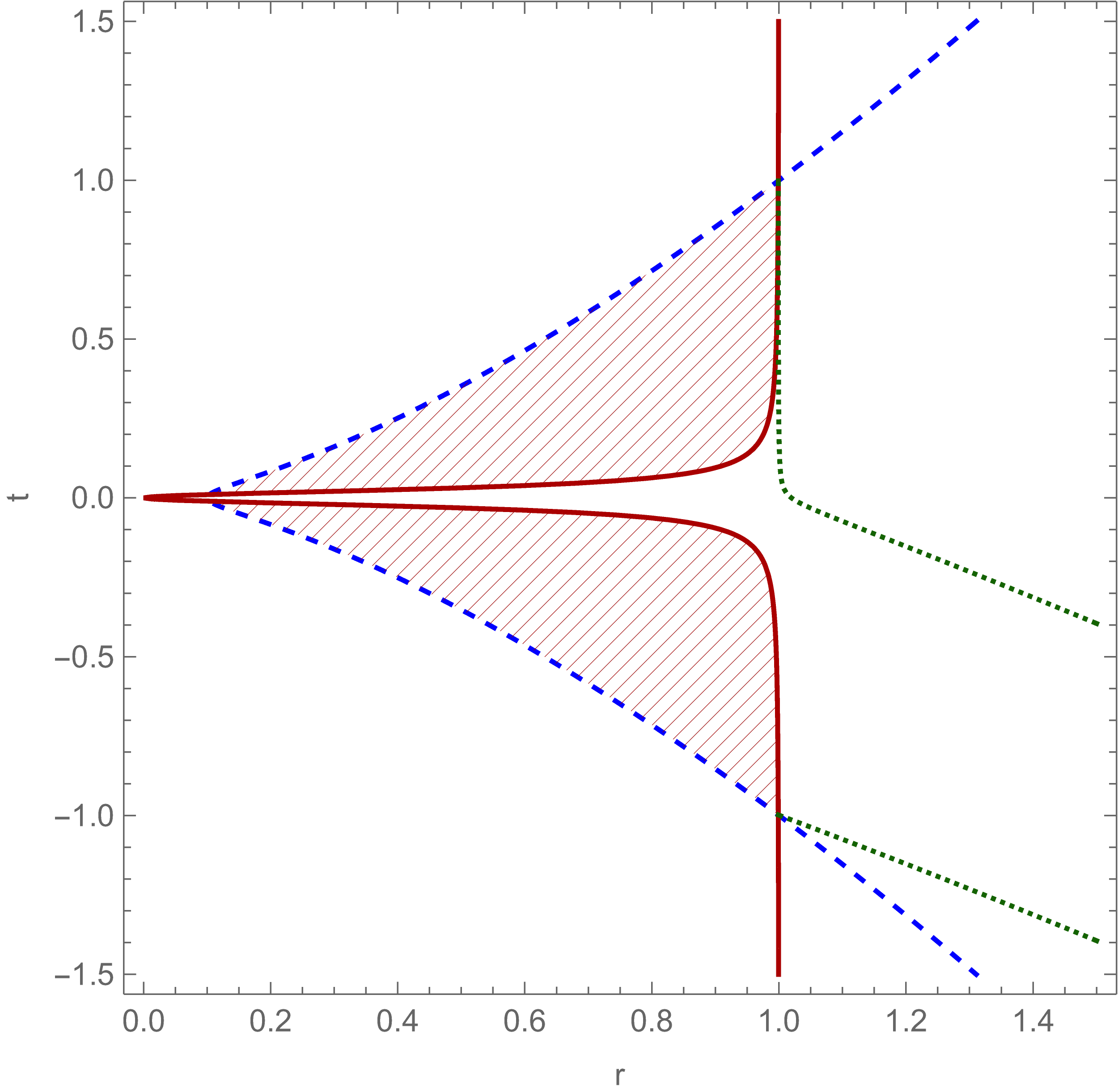}
		\caption{$\frac{R_0}{2M_0}=0.1$}
	\end{subfigure}
	\begin{subfigure}{0.45\textwidth}
		\centering
		\includegraphics[width=\textwidth]{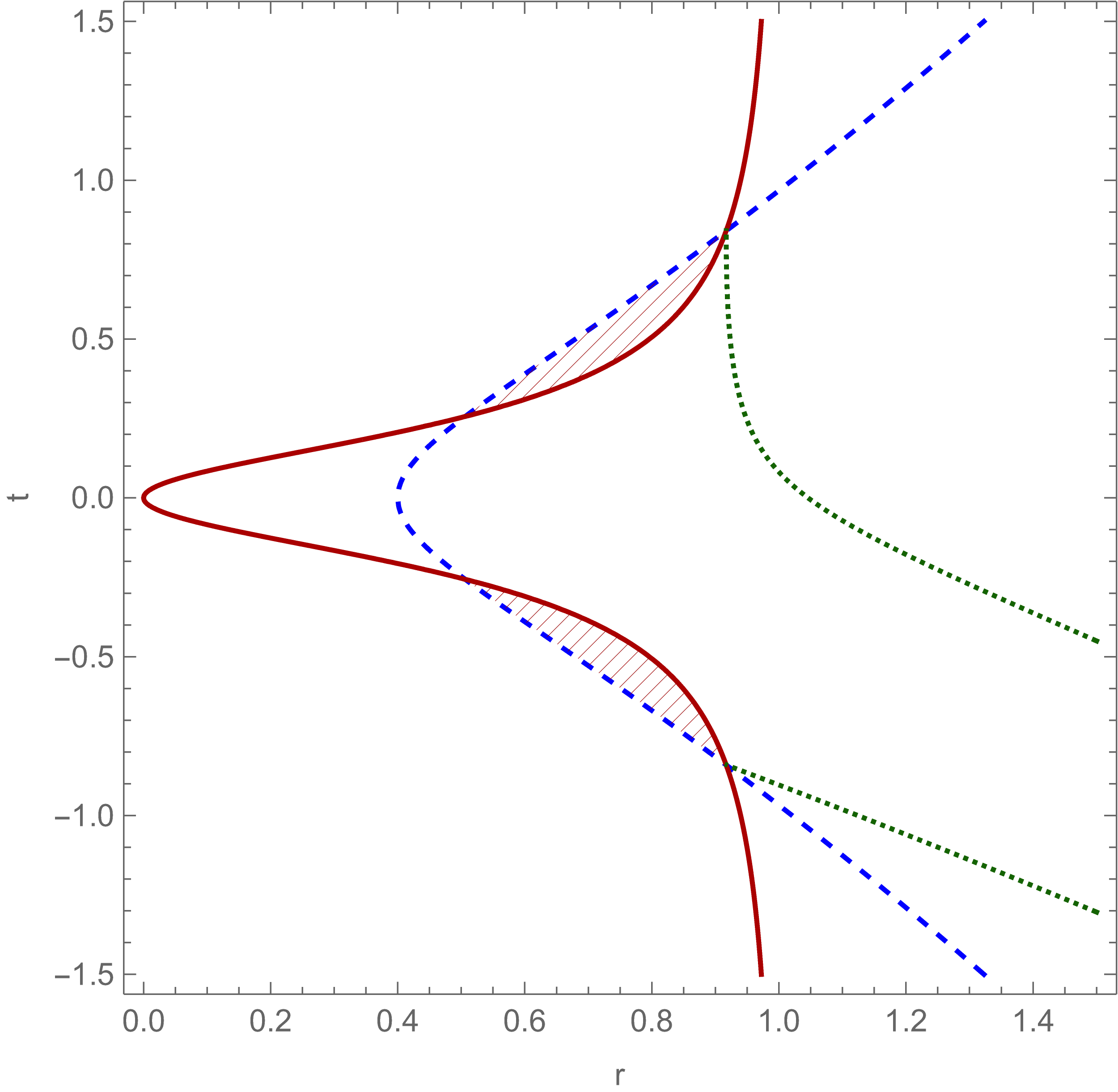}
		\caption{$\frac{R_0}{2M_0}=0.4$}
	\end{subfigure}
	\caption{The rescaled quantum corrected trajectory $R_S/2M_0$, see Eq.\ \eqref{eq:specific_trajectory}, (dashed blue line) as compared to the horizon (full red line) and the two lightrays following Eq.\ \eqref{eq:lightrays} (dotted green lines) for different values of $R_0/2M_0$. Hatching denotes the (anti-)trapped regions in the exterior.}
	\label{fig:horizons_with_light}
\end{figure}

Fig.\ \ref{fig:horizons_with_light} illustrates how different values of $R_0/2M_0$ influence the two lightrays. As is apparent, increasing $R_0/2M_0$ does allow the lightray emitted during collapse to escape to infinity earlier, but not early enough to outweigh the fact that the second lightray is also emitted earlier.

Unfortunately this lifetime is much too short to agree with astrophysical observations for any value of $R_0/2M_0$. It is notable that our approximate result $\Delta \tau_\text{ext}\propto M_0$ was found before in several different approaches to bouncing collapse, see e.g.\ \cite{AmbrusHajicekLifetime,ChristodoulouLifetime,ChristodoulouLifetime2,BarceloBounce2}.

To find $\Delta\tau_\text{ext} \approx \frac{4 M_0}{3}$ in the limit $M_0\ll R_0$, we have only used that the exterior approaches the quasi-classical black hole to white hole transition described above. It seems therefore plausible that this result also applies to other choices of $\mathcal{F}$, and more generally to other equations of motion than Eq.\ \eqref{eq:specific_eom}. The result $\Delta\tau_\text{ext} \propto M_0$ appears to be quite generic, and a significantly longer lifetime can only emerge for special cases.

Note that this conclusion can be circumvented when one softens the matching conditions to allow the collapsing body's surface to carry energy. See for an example Ref.\ \cite{WilsonQuantumOS}, where during collapse the interior is smoothly matched to an exterior, but after the bounce the reexpansion is accompanied by a shockwave. It was then found that the lifetime is proportional to $M_0^2$.

It has been proposed e.g.\ in Ref.\ \cite{BarceloBounce2} that Eardley’s white hole instability could alleviate the lifetime problem, either by prolonging the lifetime or by introducing a new equilibrium configuration at the end of bouncing collapse. Eardley's instability, discussed among others in Refs.\ \cite{EardleyInstability,BlauInstability,BarrabesInstability,OriInstability,LakeInstability,LakeInstability2}, can be summarized as follows: matter accreted onto a white hole can never pass the horizon, but only asymptotically approach it. If enough matter is accumulated close enough to the horizon such that the combined object, white hole and accreted matter, is smaller than twice its total mass, a trapping region forms just outside of the white hole. It thus effectively turns into a black hole. This phenomenon has been explored mostly for null dust as accreting matter.

In light of our discussion above that lightrays in the antitrapping region can escape the horizon due to its outward expansion, it becomes clear that whether or not Eardley's instability plays a role depends on our two timescales $\Delta\tau_{\text{trans}}$ and $\Delta\tau_{\text{hor}}$. The lightrays cannot accumulate at the horizon during its transition, they are rather swallowed up by its outward expansion until they hit the reexpanding dust cloud, see Fig.\ \ref{fig:instability}. As a result, for the effect to come into play one needs to maximize the time the exterior white hole horizon is close to stationary: $\Delta\tau_{\text{hor}}$ needs to be large and $\Delta\tau_{\text{trans}}$ small. 

It seems thus unlikely that Eardley's instability can be used to increase our $\Delta\tau_{\text{hor}}$, since this timescale already needs to be comparatively large for the instability to come into effect. Note that a similar point was made in Ref.\ \cite{BarceloBounce1}, but there the authors used it to argue for small $\Delta\tau_{\text{hor}}$ to make the bouncing scenario robust against white hole instabilities.

%%%%%%%%%%%%%%%%%%%%%%%%%%%%%%%%%%%%%%%%%%%%%%%%%%%%%%%%%%%%%%%%%%%%%%%%%

\section{Conclusions} \label{ch:conclusions}

In this article we have investigated possible exterior geometries to a quantum corrected bouncing dust cloud. We have demonstrated a straightforward way to construct such exteriors via an LTB-like construction, and we have discussed two particular examples.

The first of these was a static exterior. Under some mild assumptions about the quantum corrected equation of motion of the dust cloud we have shown that the causal structure of this exterior necessarily needs to be non-trivial: the dust cloud cannot reexpand towards the same asymptotic infinity it started its collapse from, it bounces into a different universe. As an example, for the specific equation of motion from Refs.\ \cite{MeLTB,MeQuantumOS} the causal structure of the static exterior matches that of a Reissner-Nordstr\"om black hole. Conceptually this is somewhat unsatisfying, since it would make the bounce unobservable.

To circumvent this problem one has to consider dynamic exteriors. The particular case we have looked at is a time-dependent-mass exterior: a generalization of a Schwarzschild black hole in Painlev\'{e}-Gullstrand form where the mass varies with comoving time. This leads to the position of the horizon varying with time: it emerges from the dust cloud during the collapse as in the classical case but then shrinks back into the cloud before the bounce, and reemerges from it afterwards. In this way the horizon transitions from trapping to antitrapping.

We have introduced two relevant timescales characterizing this process: $\Delta\tau_{\text{trans}}$ tells us how long the transition described above takes, and $\Delta\tau_{\text{ext}}$ determines how long there are horizons in the exterior as seen by a far away stationary observer. Especially the second one, the black hole lifetime, is of great importance for comparison with observations. The method to compute it we employed here might be useful for future investigations. It circumvents the absence of a timelike Killing vector in the dynamic exteriors we discuss, which is usually used to characterize the observer relevant for the lifetime.

For the bouncing collapse from Refs.\ \cite{MeLTB,MeQuantumOS} both of the aforementioned times have an upper bound proportional to the initial mass of the dust cloud. This is reasonable for $\Delta\tau_{\text{trans}}$ but is much too short for $\Delta\tau_{\text{ext}}$. In the light of these results we cannot claim that our time-dependent-mass exterior describes consistent bounces in an astrophysical setting. We nevertheless believe that our investigation here is useful, because it informs a more systematic search for more reasonable such candidates in the future.

For example, we have found indications that this result for $\Delta\tau_{\text{ext}}$ should also hold for other dynamic exteriors with a similar behavior of the horizon, at least when the minimal radius of the bounce is much smaller than the initial mass. This indicates that a more reasonable exterior needs to have a more complicated horizon structure. In particular, such an exterior needs to have a different quasi-classical limit than our time-dependent-mass solution.

Furthermore one can investigate other avenues to prolong the lifetime. We have already briefly touched on Eardley's white hole instability in Sec.\ \ref{sec:chapter_5B}: it can only significantly influence the bouncing scenario when the white hole horizon is present and approximately stationary long enough for matter to accrete, and hence requires an already large lifetime and short transition time. A further option to explore is Hawking radiation. As a starting point one could compute particle creation in the time-dependent-mass background, and discuss what impact backreaction could have. 

%%%%%%%%%%%%%%%%%%%%%%%%%%%%%%%%%%%%%%%%%%%%%%%%%%%%%%%%%

\section*{Acknowledgments}

The author would like to thank Claus Kiefer and Daniele Malafarina for helpful discussions and useful feedback on the manuscript.

\appendix

%%%%%%%%%%%%%%%%%%%%%%%%%%%%%%%%%%%%%%%%%%%%%%%%%%%%%%%%%%%%%%%%%%%%%%%%%

\section{Matching conditions}\label{ch:chapter_2}

Here we want to demonstrate that the exteriors constructed in Sec.\ \ref{sec:chapter_4A} given by \eqref{eq:line_element_general} are indeed matched smoothly to the dust cloud's interior. For clarity we reproduce here the line element for this interior \eqref{eq:line_element_int},
\begin{equation}
ds^2_-=-d\tau^2+R_S^2(\tau)\left(\frac{d\rho^2}{1-k_S \rho^2} +\rho^2 d\Omega^2 \right) .
\end{equation} 
The surface of the collapsing body will be the matching surface. We denote it by $\Sigma$, and on it we use $\tau$ as well as the two angular coordinates $\theta$ and $\phi$ as a coordinate frame.
%We want to keep the freedom to have our matching surface, denoted by $\Sigma$, not coincide with the surface of the collapsing body. We hence allow $\Sigma$ to vary with proper time inside of the collapsing body and let it be given by $\rho=\overline{\rho}(\tau)$, where $0<\overline{\rho}\leq1$. On $\Sigma$ we further use $\tau$ as well as the two angular coordinates $\theta$ and $\phi$ as a coordinate frame.

First we compute here the matching conditions between the interior and a generic spherically symmetric spacetime given by
\begin{equation}
ds^2_+=-e^{h(R,T)}\,f(R,T) \,dT^2 + \frac{dR^2}{f(R,T)} + R^2 d\Omega^2 . \label{eq:metric_spherical}
\end{equation} 
Later we will bring the line element \eqref{eq:line_element_general} into this form and show that it fulfills the matching conditions.

In the exterior the matching surface $\Sigma$ can be described by $R=\overline{R}(T)$. On $\Sigma$ we can further impose a coordinate transformation to the coordinates on $\Sigma$ by setting $T= \overline{T}(\tau)$. The at this point arbitrary function $\overline{T}(\tau)$ will be determined by the matching procedure. 

For the matching procedure we follow Ref.\ \cite{PoissonRelativistsToolkit}. The first matching condition we impose is
\begin{equation}
\left. ds^2_-\right|_\Sigma = \left. ds^2_+\right|_\Sigma .
\end{equation}
Matching the angular components of the metrics on $\Sigma$ leads to
\begin{equation}
\overline{R}(\overline{T}(\tau)) = R_S(\tau) . \label{eq:matching_1}
\end{equation}
%\begin{equation}
%	\overline{R}(\overline{T}(\tau)) = R_S(\tau)\, \overline{\rho}(\tau) . \label{eq:matching_1}
%\end{equation}
With this one can write the remaining condition from the $\tau\tau$ components of the metrics as 
\begin{equation}
1 = e^{\overline{h}(\tau)}\,\overline{f}(\tau)\, \dot{\overline{T}}^2(\tau) - \frac{\dot{R}_S^2(\tau)}{\overline{f}(\tau)} , \label{eq:matching_2} 
\end{equation}
%\begin{equation}
%1 - \frac{R_S^2(\tau)\dot{\overline{\rho}}^2(\tau)}{1 - k_S \overline{\rho}^2(\tau)} = e^{\overline{h}(\tau)}\,\overline{f}(\tau)\, \dot{\overline{T}}^2(\tau) - \frac{\dot{\overline{R}}^2(\tau)}{\overline{f}(\tau)} , \label{eq:matching_2} 
%\end{equation}
where a dot denotes a derivative with regard to $\tau$. An overline over a functions means it is evaluated on $\Sigma$, for a function $q(R,T)$ we have $\overline{q}=q(\overline{R}(\overline{T}(\tau)),\overline{T}(\tau))$. In the following we will for clarity not explicitly denote the $\tau$ dependency of quantities on $\Sigma$. With the exterior metric, the equation of motion $\dot{R}_S$ %and the trajectory of the matching surface $\overline{\rho}(\tau)$ 
fixed, this determines $\overline{T}(\tau)$.

For the second matching condition we need to compute the extrinsic curvature of $\Sigma$ from both sides. To this end we note that the unit normal co-vectors to $\Sigma$ in the interior and exterior are
\begin{align}
n^-_\mu dx_-^\mu&=\frac{R_S(\tau) d\rho  }{\sqrt{ 1-k_S }} ,\\
n^+_\mu dx_+^\mu&=\frac{ \sqrt{\left|f(R,T)\right| }\left( dR -  \frac{d\overline{R}}{dT} dT\right) }{\sqrt{\left| f^2(R,T) - e^{-h(R,T)} \left( \frac{d\overline{R}}{dT}\right)^2  \right| }} ,
\end{align}
%\begin{align}
%n^-_\mu dx_-^\mu&=\frac{R_S(\tau) \left(d\rho - \dot{\overline{\rho}}(\tau)d\tau \right) }{\sqrt{\left| 1-k_S \rho^2-R_S^2(\tau) \dot{\overline{\rho}}^2(\tau)\right| }} ,\\
%n^+_\mu dx_+^\mu&=\frac{ \sqrt{\left|f(R,T)\right| }\left( dR -  \frac{d\overline{R}}{dT} dT\right) }{\sqrt{\left| f^2(R,T) - e^{-h(R,T)} \left( \frac{d\overline{R}}{dT}\right)^2  \right| }} ,
%\end{align}
where $x_\pm^\mu$ denote the coordinates in the exterior and interior, respectively. With $y^a$ being our coordinates on $\Sigma$, the extrinsic curvature tensors are then
\begin{align}
K^\pm_{ab}&=\frac{\partial x_\pm^\mu}{\partial y^a}\frac{\partial x_\pm^\nu}{\partial y^b}\nabla_\mu n^\pm_\nu = \frac{\partial x_\pm^\mu}{\partial y^a}\frac{\partial x_\pm^\nu}{\partial y^b}\left( \partial_\mu n^\pm_\nu - \Gamma^\sigma_{\mu\nu} n^\pm_\sigma \right)\\
&= -n^\pm_\sigma\left( \frac{\partial^2 x_\pm^\sigma}{\partial y^b \partial y^a} + {\Gamma_{\!\!\pm}}^\sigma_{\mu\nu} \frac{\partial x_\pm^\nu}{\partial y^b} \frac{\partial x_\pm^\mu}{\partial y^a}\right) ,
\end{align}
where $ {\Gamma_{\!\!\pm}}^\sigma_{\mu\nu}$ are the Christoffel symbols with regard to the interior and exterior metric, respectively, and we have used that by definition the normal vectors are orthogonal to the projectors onto $\Sigma$,
\begin{equation}
n^\pm_\mu \frac{\partial x_\pm^\mu}{\partial y^a} = 0 .
\end{equation}
The non-vanishing components of the extrinsic curvature tensor in the interior are then
\begin{align}
K^-_{\theta\theta}&= R_S \sqrt{ 1-k_S } ,\\
K^-_{\phi\phi}&=K^-_{\theta\theta}\,\sin^2\theta ,
\end{align}
%\begin{align}
%K^-_{\tau\tau} &=- N^-\left( R_S \ddot{\overline{\rho}} +  2 \dot{R}_S \dot{\overline{\rho}} + R_S \dot{\overline{\rho}}^2 \frac{k_S \overline{\rho} - R_S \dot{R}_S \dot{\overline{\rho}}}{1-k_S \overline{\rho}^2} \right) , \\
%K^-_{\theta\theta}&=N^- R_S \,\overline{\rho}\left(1-k_S \overline{\rho}^2+R_S \dot{R}_S \overline{\rho} \, \dot{\overline{\rho}} \right) ,\\
%K^-_{\phi\phi}&=K^-_{\theta\theta}\,\sin^2\theta ,
%\end{align}
%where
%\begin{equation}
%	N^-=\frac{1}{\sqrt{\left| 1-k_S \overline{\rho}^2-R_S^2 \dot{\overline{\rho}}^2\right| }}
%\end{equation}
In the exterior we consider first the angular components
\begin{equation}
K^+_{\theta\theta} = N^+ R_S \,\overline{f} \quad\text{and}\quad K^+_{\phi\phi}=K^+_{\theta\theta}\,\sin^2\theta ,
\end{equation}
where 
\begin{equation}
N^+ = \sqrt{\left|\frac{ \overline{f}\,\dot{\overline{T}}^2 }{\overline{f}^2 \dot{\overline{T}}^2 - e^{-\overline{h}} \dot{R}_S^2 }\right| }  .
\end{equation}
We have also replaced $\frac{d\overline{R}}{dT}$ with the help of the identity $\dot{R}_S=\frac{d\overline{R}}{dT}\,\dot{\overline{T}}$ following from Eq.\ \eqref{eq:matching_1}.

We assume that quantum corrections do not lead to a distinguished distributional contribution to the energy momentum tensor on the surface of the collapsing body, meaning that we can impose $K^+_{ab}=K^-_{ab}$ as further matching conditions. Matching the angular components of $K^\pm_{ab}$ gives then with Eq.\ \eqref{eq:matching_1} the condition
\begin{equation}
N^+ \,\overline{f} =\sqrt{1-k_S} , \label{eq:matching_fiducial}
\end{equation}
%\begin{equation}
%N^+ \,\overline{f} = N^- \left(1-k_S \overline{\rho}^2+R_S \dot{R}_S \overline{\rho} \, \dot{\overline{\rho}} \right) , %\label{eq:matching_fiducial}
%\end{equation}
which can be simplified with Eq.\ \eqref{eq:matching_2} to
\begin{equation}
\dot{\overline{T}}^2  = \left(1-k_S \right) \frac{e^{-\overline{h}}}{\overline{f}^2} . \label{eq:matching_3}
\end{equation}
%\begin{equation}
%\dot{\overline{T}}^2  = \frac{ \left(1-k_S \overline{\rho}^2+R_S \dot{R}_S \overline{\rho} \, \dot{\overline{\rho}} \right)^2}{1-k_S \overline{\rho}^2} \,\frac{e^{-\overline{h}}}{\overline{f}^2} . \label{eq:matching_3}
%\end{equation}
Plugging this into Eq.\ \eqref{eq:matching_2} and using condition \eqref{eq:matching_1} differentiated with respect to $\tau$ we find
\begin{equation}
\dot{R}_S^2 = 1-k_S - \overline{f} . \label{eq:matching_4}
\end{equation}
%\begin{equation}
%\dot{R}_S^2 \,\overline{\rho}^2 = 1-k_S \overline{\rho}^2 - \overline{f} . \label{eq:matching_4}
%\end{equation}
As a consistency check it is straightforward to see that for a Schwarzschild exterior %and $\overline{\rho}=1$ 
this matching condition gives us the first Friedmann equation with dust as matter, where one can identify $M=\frac{4\pi}{3}\epsilon R_S^3$ and $\epsilon$ is the dust density. 

With the help of the conditions \eqref{eq:matching_1}, \eqref{eq:matching_fiducial}, \eqref{eq:matching_3} and \eqref{eq:matching_4} the last non-vanishing component of $K^+_{ab}$, and with it the last matching condition, can be expressed as
%\begin{widetext}
%\begin{multline}
%	K^+_{\tau \tau} - K^-_{\tau \tau} = -\frac{N^+}{2}\frac{1-k_S \overline{\rho}^2 - R_S^2 \dot{\overline{\rho}}^2}{1-k_S \overline{\rho}^2}\left\{ \left[1 + \frac{(1-k_S \overline{\rho}^2 - R_S^2 \dot{\overline{\rho}}^2) \dot{R}_S^2 \overline{\rho}^2 }{(1-k_S \overline{\rho}^2+R_S \dot{R}_S \overline{\rho} \, \dot{\overline{\rho}} )^2} \right] 2\ddot{R}_S \overline{\rho} + \left[1 + \frac{(1-k_S \overline{\rho}^2) \dot{\overline{R}}^2 }{(1-k_S \overline{\rho}^2+R_S \dot{R}_S \overline{\rho} \, \dot{\overline{\rho}} )^2} \right] \overline{\frac{\partial f}{\partial R}} \right. \\ \left.  + \overline{f}\, \overline{\frac{\partial h}{\partial R} } + \frac{2\overline{\rho}\, \dot{\overline{\rho}} (\dot{R}_S^2 + k_S) }{(1-k_S \overline{\rho}^2+R_S \dot{R}_S \overline{\rho} \, \dot{\overline{\rho}} )^2} \left[ \dot{\overline{R}}(1-k_S\overline{\rho}^2) + \dot{R}_S \overline{\rho} (1-k_S \overline{\rho}^2+R_S \dot{R}_S \overline{\rho} \, \dot{\overline{\rho}} ) \right]  \right\} = 0. \label{eq:matching_5}
%\end{multline}
%\end{widetext}

%This condition simplifies immensely for when the matching surface coincides with the collapsing matter's surface, $\overline{\rho}=1$,
\begin{equation}
-\frac{N^+}{2} \left[ \frac{1-k_S + \dot{R}_S^2 }{1-k_S} \left(  2\ddot{R}_S +  \overline{\frac{\partial f}{\partial R}} \right)  + \overline{f}\, \overline{\frac{\partial h}{\partial R} }  \right] = 0. \label{eq:matching_last}
\end{equation}
For a Schwarzschild exterior this is equivalent to the second Friedmann equation for dust. 

In summary we can say that the matching conditions here take the form of Eqs. \eqref{eq:matching_1}, \eqref{eq:matching_3}, \eqref{eq:matching_4} and \eqref{eq:matching_last}.

We now want to demonstrate that the metric \eqref{eq:line_element_general} fulfills these conditions. To this end we first have to bring it into the form \eqref{eq:metric_spherical}. In analogy with Schwarzschild this can be achieved by a coordinate transformation $\tau(T,R)$, where
\begin{equation}
\frac{\partial \tau}{\partial R} = -\frac{\mathcal{F}}{1-k_S-\mathcal{F}^2} . \label{eq:coord_trafo}
\end{equation}
This gives us
\begin{multline}
ds^2 = - \frac{1 - k_S - \mathcal{F}^2}{1 - k_S}\left(\frac{\partial \tau}{\partial T} \right)^2 dT^2 \\+ \frac{dR^2}{1 - k_S - \mathcal{F}^2} + R^2\, d\Omega^2 , \label{eq:line_element_diagonal}
\end{multline}
from which we can identify
\begin{align}
f &= 1 - k_S - \mathcal{F}^2,\\
e^h &= \frac{1}{1 - k_S} \left(\frac{\partial \tau}{\partial T} \right)^2 .
\end{align}

Now we can check the matching conditions one by one. The matching surface $\Sigma$ is defined by $\rho = 1$, or according to Eq.\ \eqref{eq:ansatz1} equivalently by $R = R_S(\tau)$. Together with our previous characterization of $\Sigma$ in diagonal coordinates by $R=\overline{R}(T)$ this directly implies matching condition \eqref{eq:matching_1}. 

On the matching surface we further have
\begin{align}
\overline{\mathcal{F}}=\mathcal{F}(R_S,R_S)=\left. \frac{\partial R}{\partial \tau} \right|_{\rho=1} =\dot{R}_S,
\end{align}
as already mentioned in Sec.\ \ref{sec:chapter_4A}. With this we can show that the matching condition \eqref{eq:matching_4} is fulfilled,
\begin{equation}
\dot{R}_S^2 = \overline{\mathcal{F}}^2 = 1-k_S - \overline{f}. \label{eq:trajectory_from_matching_1}
\end{equation}

Inverting the coordinate transformation given by Eq.\ \eqref{eq:coord_trafo} we find
\begin{align}
\frac{\partial T}{\partial \tau} &= \left( \frac{\partial \tau}{\partial T}\right)^{-1},\\
\frac{\partial T}{\partial R} &= \frac{\mathcal{F}}{\left( 1-k_S-\mathcal{F}^2\right) \frac{\partial \tau}{\partial T} }.
\end{align}
On $\Sigma$ we thus have, using Eq.\ \eqref{eq:fiducial_identity},
\begin{align}
\dot{\overline{T}} &= \overline{\frac{\partial T}{\partial \tau}} + \overline{\frac{\partial T}{\partial R}} \dot{R}_S = \left( \overline{\frac{\partial \tau}{\partial T}}\right)^{-1} \frac{1 - k_S}{1-k_S-\overline{\mathcal{F}}^2} \label{eq:trajectory_from_matching_2} \\&= \sqrt{1-k_S} \frac{e^{-\frac{\overline{h}}{2}}}{\overline{f}},
\end{align}
which is identical to the matching condition \eqref{eq:matching_3}.

The last condition to check is Eq.\ \eqref{eq:matching_last}. To this end we first compute
\begin{align}
2\ddot{R}_S + \overline{\frac{\partial f}{\partial R}} &= -\overline{\frac{\partial f}{\partial T}}\, \frac{\dot{\overline{T}}}{\dot{R}_S} = 2 \overline{\mathcal{F}} \overline{\frac{\partial \mathcal{F}}{\partial R_S}} \frac{1 - k_S}{1-k_S-\overline{\mathcal{F}}^2} ,\\
\overline{f}\overline{\frac{\partial h}{\partial R}} &= 2 \overline{f} \left( \overline{\frac{\partial \tau}{\partial T}}\right)^{-1} \overline{\frac{\partial^2 \tau}{\partial T\partial R}}\\
&=- 2 \overline{\mathcal{F}} \overline{\frac{\partial\mathcal{F}}{\partial R_S}}  \frac{1-k_S+\overline{\mathcal{F}}^2}{1-k_S-\overline{\mathcal{F}}^2  }  .
\end{align}
Plugging this into Eq.\ \eqref{eq:matching_last} shows that this last matching condition is also fulfilled. The metric given in Eq.\ \eqref{eq:line_element_general} or equivalently Eq.\ \eqref{eq:line_element_diagonal} thus can be smoothly matched to our quantum corrected FLRW metric.

\section{Penrose diagram}\label{app:A}

Here we want to discuss the construction of the Penrose diagram in Fig.\ \ref{fig:penrose} to the static exterior following Ref.\ \cite{SchindlerDiagram}. To start with we want to note that we technically do not draw a Penrose diagram according to its definition in Ref.\ \cite{SchindlerDiagram}, since we do not explicitly construct a global coordinate frame for the maximal extension of the spacetime. We rather settle for what is called in Ref.\ \cite{SchindlerDiagram} a block diagram, which lacks the coordinate frame but still illustrates the global causal structure of the spacetime. In an abuse of terminology we will still refer to it as a Penrose diagram, as is common in the literature.

Central to what we will do in the following is the introduction of double null coordinates $U=T-R_*$ and $V=T+R_*$, where
\begin{equation}
	R_*(R)=\int\frac{dR}{f(R)},
\end{equation}
where $f(R)$ is given by Eq.\ \eqref{eq:static_f}. The behavior of $R_*$ and in particular its divergences determine the structure of the Penrose diagram. Recall that the roots of $f(R)$ determine the position of horizons in the spacetime. Since we have assumed that these roots are simple, we can say that $R_*$ diverges there and behaves monotonically in between. In these intervals between two roots $U$ and $V$ span thus the whole of $\mathds{R}^2$, and hence these intervals can be identified with full diamonds in the block diagram. In this way, all intervals of $R$, between roots of $f(R)$ and also $R\to0$ and $R\to\infty$, correspond to blocks in the diagram, with their shape determined by the behavior of $R_*$ at the interval's boundaries.

\begin{figure}[h]
	\begin{subfigure}{0.45\textwidth}
		\centering
		%\begin{tikzpicture}
		%\draw (-1,0) -- (0,1) node [midway, above, sloped] (hor1) {$R_\text{inner}$};
		%\draw (0,1) -- (1,0) node [midway, above, sloped] (hor2) {$R_\text{inner}$};
		%\draw (1,0) -- (0,-1) node [midway, below, sloped] (hor3) {$R_\text{outer}$};
		%\draw (0,-1) -- (-1,0) node [midway, below, sloped] (hor4) {$R_\text{outer}$};
		%\draw (0,0) node (label) {trapped};
		%\draw (-1+3,0) -- (0+3,1) node [midway, above, sloped] (hor1) {$R_\text{outer}$};
		%\draw (0+3,1) -- (1+3,0) node [midway, above, sloped] (hor2) {$R_\text{outer}$};
		%\draw (1+3,0) -- (0+3,-1) node [midway, below, sloped] (hor3) {$R_\text{inner}$};
		%\draw (0+3,-1) -- (-1+3,0) node [midway, below, sloped] (hor4) {$R_\text{inner}$};
		%\draw (0+3,0) node [align=center] (label) {anti-\\trapped};
		%\end{tikzpicture}
		\includegraphics{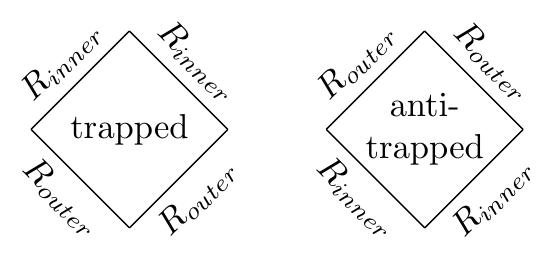}
		\caption{$R_\text{inner}<R<R_\text{outer}$}
		\label{fig:block1}
	\end{subfigure}
	\begin{subfigure}{0.45\textwidth}
		\centering
		%\begin{tikzpicture}
		%\draw (0,1) -- (1,0) node [midway, above, sloped] (hor2) {$R_\text{outer}$};
		%\draw (1,0) -- (0,-1) node [midway, below, sloped] (hor3) {$R_\text{outer}$};
		%\draw[very thick] (0,-1) -- (-1,0)node [at end, left] (node1) {$R\to\infty$} -- (0,1);
		%\draw (-1+3,0) -- (0+3,1) node [midway, above, sloped] (hor1) {$R_\text{outer}$};
		%\draw[very thick] (0+3,1) -- (1+3,0)node [at end, right] (node1) {$R\to\infty$} -- (0+3,-1);
		%\draw (0+3,-1) -- (-1+3,0) node [midway, below, sloped] (hor4) {$R_\text{outer}$};
		%\end{tikzpicture}
		\includegraphics{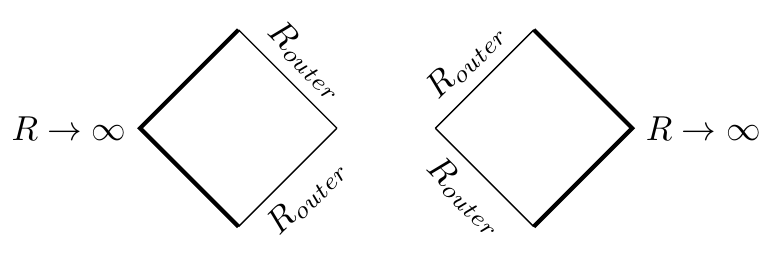}
		\caption{$R_\text{outer}<R<\infty$}
		\label{fig:block2}
	\end{subfigure}
	\begin{subfigure}{0.45\textwidth}
		\centering
		%\begin{tikzpicture}
		%\draw (-1,0) -- (0,1) node [midway, above, sloped] (hor2) {$R_\text{inner}$};
		%\draw[snake=coil,segment aspect=0, segment length=8.75] (0,1) -- (0,-1) node [midway, above, sloped] (hor2) {$R\to0$};
		%\draw (0,-1) -- (-1,0) node [midway, below, sloped] (hor2) {$R_\text{inner}$};
		%\draw (0+3,1) -- (1+3,0) node [midway, above, sloped] (hor2) {$R_\text{inner}$};
		%\draw (1+3,0) -- (0+3,-1) node [midway, below, sloped] (hor2) {$R_\text{inner}$};
		%\draw[snake=coil,segment aspect=0, segment length=8.75, mirror snake] (0+3,1) -- (0+3,-1) node [midway, below, sloped] (hor2) {$R\to0$};
		%\end{tikzpicture}
		\includegraphics{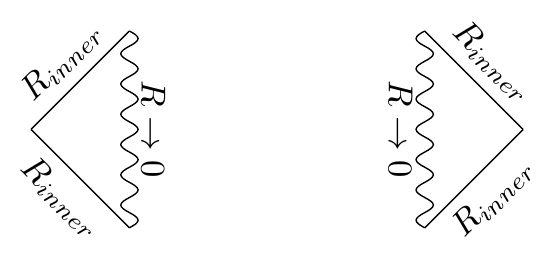}
		\caption{$0<R<R_\text{inner}$}
		\label{fig:block3}
	\end{subfigure}
	\caption{The different blocks in the Penrose diagram in all orientations. Thick lines denote null infinities, thin lines horizons, and wavy lines singularities.}
\end{figure}

In our case $R_*(R)$ can be given in analytic form. Factorizing $f(R)$ according to its roots, and decomposing the fraction into a sum we find
\begin{equation}
	R_*(R)=R-\sum_{R_i:f(R_i)=0} \frac{\ln|R-R_i|}{f'(R_i)}.
\end{equation}
We restrict ourselves here to the case $2^\frac{5}{3}R_0<3 M_0$, for which $f(R)$ has two real roots $R_\text{inner}<R_\text{outer}$. Since $f(R\to0)\to\infty$ and $f(R\to\infty)\to1$, we know that $f'(R_\text{inner})<0$ and $f'(R_\text{outer})>0$. We have then $R_*(R\to R_\text{inner})\to \infty$ and $R_*(R\to R_\text{outer})\to -\infty$, and hence this interval in $R$ is associated with a full diamond in the block diagram bounded by the horizons $R_\text{inner}$ and $R_\text{outer}$. Since we further have $f(R)<0$ there, this region is trapped or antitrapped and the lower two edges of the diamond are given by the horizon $R_\text{inner}$ and the upper ones by $R_\text{outer}$, or the other way around. In Fig.\ \ref{fig:block1} we draw this block in its two orientations. 

With $R_*(R\to\infty)\to\infty$ we can analogously proceed for $R_\text{outer}<R<\infty$. Since this region is untrapped, the diamond is bounded by the horizon on the left and the null infinities on the right, or the other way around. We illustrate how this block looks in Fig.\ \ref{fig:block2}.

In the last interval, $0<R<R_\text{inner}$, we find that $R_*(R\to0)\to\text{const.}$ and hence $U$ and $V$ cannot fill the whole diamond; $V-U$ is bounded from above, and the corresponding block is only half of a diamond terminating in the singularity $R\to0$, see Fig.\ \ref{fig:block3}.

To find our Penrose diagram in Fig.\ \ref{fig:penrose}, we now need to fit these blocks together at matching horizons until there is no horizon left to be matched. This construction is for our case unique, and leads to a maximally extended spacetime that is formed by an infinite chain of blocks, similar to a Reissner-Nordstr\"om black hole.

%%%%%%%%%%%%%%%%%%%%%%%%%%%%%%%%%%%%%%%%%%%%%%%%%%%%%%%%%%%%%%%%%%%%%%%%%

%\section{appendix B}\label{app:B}

%%%%%%%%%%%%%%%%%%%%%%%%%%%%%%%%%%%%%%%%%%%%%%%%%%%%%%%%%%%%%%%%%%

%bibtex, remove before submission
%\bibliography{ext_references}

\begin{thebibliography}{57}%
	\makeatletter
	\providecommand \@ifxundefined [1]{%
		\@ifx{#1\undefined}
	}%
	\providecommand \@ifnum [1]{%
		\ifnum #1\expandafter \@firstoftwo
		\else \expandafter \@secondoftwo
		\fi
	}%
	\providecommand \@ifx [1]{%
		\ifx #1\expandafter \@firstoftwo
		\else \expandafter \@secondoftwo
		\fi
	}%
	\providecommand \natexlab [1]{#1}%
	\providecommand \enquote  [1]{``#1''}%
	\providecommand \bibnamefont  [1]{#1}%
	\providecommand \bibfnamefont [1]{#1}%
	\providecommand \citenamefont [1]{#1}%
	\providecommand \href@noop [0]{\@secondoftwo}%
	\providecommand \href [0]{\begingroup \@sanitize@url \@href}%
	\providecommand \@href[1]{\@@startlink{#1}\@@href}%
	\providecommand \@@href[1]{\endgroup#1\@@endlink}%
	\providecommand \@sanitize@url [0]{\catcode `\\12\catcode `\$12\catcode
		`\&12\catcode `\#12\catcode `\^12\catcode `\_12\catcode `\%12\relax}%
	\providecommand \@@startlink[1]{}%
	\providecommand \@@endlink[0]{}%
	\providecommand \url  [0]{\begingroup\@sanitize@url \@url }%
	\providecommand \@url [1]{\endgroup\@href {#1}{\urlprefix }}%
	\providecommand \urlprefix  [0]{URL }%
	\providecommand \Eprint [0]{\href }%
	\providecommand \doibase [0]{http://dx.doi.org/}%
	\providecommand \selectlanguage [0]{\@gobble}%
	\providecommand \bibinfo  [0]{\@secondoftwo}%
	\providecommand \bibfield  [0]{\@secondoftwo}%
	\providecommand \translation [1]{[#1]}%
	\providecommand \BibitemOpen [0]{}%
	\providecommand \bibitemStop [0]{}%
	\providecommand \bibitemNoStop [0]{.\EOS\space}%
	\providecommand \EOS [0]{\spacefactor3000\relax}%
	\providecommand \BibitemShut  [1]{\csname bibitem#1\endcsname}%
	\let\auto@bib@innerbib\@empty
	%</preamble>
	\bibitem [{\citenamefont {Kiefer}\ and\ \citenamefont {Schmitz}(2019)}]{MeLTB}%
	\BibitemOpen
	\bibfield  {author} {\bibinfo {author} {\bibfnamefont {C.}~\bibnamefont
			{Kiefer}}\ and\ \bibinfo {author} {\bibfnamefont {T.}~\bibnamefont
			{Schmitz}},\ }\href {\doibase 10.1103/PhysRevD.99.126010} {\bibfield
		{journal} {\bibinfo  {journal} {Phys. Rev. D}\ }\textbf {\bibinfo {volume}
			{99}},\ \bibinfo {pages} {126010} (\bibinfo {year} {2019})}\BibitemShut
	{NoStop}%
	\bibitem [{\citenamefont {Piechocki}\ and\ \citenamefont
		{Schmitz}(2020)}]{MeQuantumOS}%
	\BibitemOpen
	\bibfield  {author} {\bibinfo {author} {\bibfnamefont {W.}~\bibnamefont
			{Piechocki}}\ and\ \bibinfo {author} {\bibfnamefont {T.}~\bibnamefont
			{Schmitz}},\ }\href {\doibase 10.1103/PhysRevD.102.046004} {\bibfield
		{journal} {\bibinfo  {journal} {Phys. Rev. D}\ }\textbf {\bibinfo {volume}
			{102}},\ \bibinfo {pages} {046004} (\bibinfo {year} {2020})}\BibitemShut
	{NoStop}%
	\bibitem [{\citenamefont {H{\'a}j{\'i}\v{c}ek}\ and\ \citenamefont
		{Kiefer}(2001{\natexlab{a}})}]{HajicekKieferNullShells}%
	\BibitemOpen
	\bibfield  {author} {\bibinfo {author} {\bibfnamefont {P.}~\bibnamefont
			{H{\'a}j{\'i}\v{c}ek}}\ and\ \bibinfo {author} {\bibfnamefont
			{C.}~\bibnamefont {Kiefer}},\ }\href@noop {} {\bibfield  {journal} {\bibinfo
			{journal} {Int. J. Mod. Phys. D}\ }\textbf {\bibinfo {volume} {10}},\
		\bibinfo {pages} {775} (\bibinfo {year} {2001}{\natexlab{a}})}\BibitemShut
	{NoStop}%
	\bibitem [{\citenamefont {Kiefer}()}]{KieferNullShellConf}%
	\BibitemOpen
	\bibfield  {author} {\bibinfo {author} {\bibfnamefont {C.}~\bibnamefont
			{Kiefer}},\ }\href@noop {} {}\Eprint {http://arxiv.org/abs/1512.08346}
	{arXiv:1512.08346 [gr-qc]} \BibitemShut {NoStop}%
	\bibitem [{\citenamefont {H{\'a}j{\'i}\v{c}ek}\ and\ \citenamefont
		{Kiefer}(2001{\natexlab{b}})}]{HajicekClassNullShells}%
	\BibitemOpen
	\bibfield  {author} {\bibinfo {author} {\bibfnamefont {P.}~\bibnamefont
			{H{\'a}j{\'i}\v{c}ek}}\ and\ \bibinfo {author} {\bibfnamefont
			{C.}~\bibnamefont {Kiefer}},\ }\href {\doibase 10.1016/S0550-3213(01)00141-9}
	{\bibfield  {journal} {\bibinfo  {journal} {Nucl. Phys. B}\ }\textbf
		{\bibinfo {volume} {603}},\ \bibinfo {pages} {531} (\bibinfo {year}
		{2001}{\natexlab{b}})}\BibitemShut {NoStop}%
	\bibitem [{\citenamefont
		{H{\'a}j{\'i}\v{c}ek}(2001)}]{HajicekQuantumNullShells}%
	\BibitemOpen
	\bibfield  {author} {\bibinfo {author} {\bibfnamefont {P.}~\bibnamefont
			{H{\'a}j{\'i}\v{c}ek}},\ }\href {\doibase 10.1016/S0550-3213(01)00140-7}
	{\bibfield  {journal} {\bibinfo  {journal} {Nucl. Phys. B}\ }\textbf
		{\bibinfo {volume} {603}},\ \bibinfo {pages} {555} (\bibinfo {year}
		{2001})}\BibitemShut {NoStop}%
	\bibitem [{\citenamefont {H{\'a}j{\'i}\v{c}ek}(2003)}]{HajicekNullShells}%
	\BibitemOpen
	\bibfield  {author} {\bibinfo {author} {\bibfnamefont {P.}~\bibnamefont
			{H{\'a}j{\'i}\v{c}ek}},\ }\enquote {\bibinfo {title} {{Quantum Theory of
				Gravitational Collapse (Lecture Notes on Quantum Conchology)}},}\ in\ \href
	{\doibase 10.1007/978-3-540-45230-0_6} {\emph {\bibinfo {booktitle} {{Quantum
					Gravity: From Theory to Experimental Search}}}},\ \bibinfo {editor} {edited
		by\ \bibinfo {editor} {\bibfnamefont {D.~J.~W.}\ \bibnamefont {Giulini}},
		\bibinfo {editor} {\bibfnamefont {C.}~\bibnamefont {Kiefer}}, \ and\ \bibinfo
		{editor} {\bibfnamefont {C.}~\bibnamefont {L{\"a}mmerzahl}}}\ (\bibinfo
	{publisher} {Springer},\ \bibinfo {address} {Berlin, Heidelberg},\ \bibinfo
	{year} {2003})\ pp.\ \bibinfo {pages} {255--299}\BibitemShut {NoStop}%
	\bibitem [{\citenamefont {Frolov}\ and\ \citenamefont
		{Vilkovisky}(1981)}]{FrolovNullShell}%
	\BibitemOpen
	\bibfield  {author} {\bibinfo {author} {\bibfnamefont {V.}~\bibnamefont
			{Frolov}}\ and\ \bibinfo {author} {\bibfnamefont {G.}~\bibnamefont
			{Vilkovisky}},\ }\href {\doibase
		https://doi.org/10.1016/0370-2693(81)90542-6} {\bibfield  {journal} {\bibinfo
			{journal} {Phys. Lett.}\ }\textbf {\bibinfo {volume} {106B}},\ \bibinfo
		{pages} {307 } (\bibinfo {year} {1981})}\BibitemShut {NoStop}%
	\bibitem [{\citenamefont {Almeida}\ \emph {et~al.}(2018)\citenamefont
		{Almeida}, \citenamefont {Bergeron}, \citenamefont {Gazeau},\ and\
		\citenamefont {Scardua}}]{AlmeidaACS}%
	\BibitemOpen
	\bibfield  {author} {\bibinfo {author} {\bibfnamefont {C.~R.}\ \bibnamefont
			{Almeida}}, \bibinfo {author} {\bibfnamefont {H.}~\bibnamefont {Bergeron}},
		\bibinfo {author} {\bibfnamefont {J.~P.}\ \bibnamefont {Gazeau}}, \ and\
		\bibinfo {author} {\bibfnamefont {A.~C.}\ \bibnamefont {Scardua}},\ }\href
	{\doibase https://doi.org/10.1016/j.aop.2018.03.010} {\bibfield  {journal}
		{\bibinfo  {journal} {Ann. Phys. (N.Y.)}\ }\textbf {\bibinfo {volume}
			{392}},\ \bibinfo {pages} {206 } (\bibinfo {year} {2018})}\BibitemShut
	{NoStop}%
	\bibitem [{\citenamefont {Bergeron}\ \emph {et~al.}(2014)\citenamefont
		{Bergeron}, \citenamefont {Dapor}, \citenamefont {Gazeau},\ and\
		\citenamefont {Ma\l{}kiewicz}}]{BergeronBounce}%
	\BibitemOpen
	\bibfield  {author} {\bibinfo {author} {\bibfnamefont {H.}~\bibnamefont
			{Bergeron}}, \bibinfo {author} {\bibfnamefont {A.}~\bibnamefont {Dapor}},
		\bibinfo {author} {\bibfnamefont {J.~P.}\ \bibnamefont {Gazeau}}, \ and\
		\bibinfo {author} {\bibfnamefont {P.}~\bibnamefont {Ma\l{}kiewicz}},\ }\href
	{\doibase 10.1103/PhysRevD.89.083522} {\bibfield  {journal} {\bibinfo
			{journal} {Phys. Rev. D}\ }\textbf {\bibinfo {volume} {89}},\ \bibinfo
		{pages} {083522} (\bibinfo {year} {2014})}\BibitemShut {NoStop}%
	\bibitem [{\citenamefont {Bergeron}\ \emph {et~al.}(2019)\citenamefont
		{Bergeron}, \citenamefont {Czuchry}, \citenamefont {Gazeau},\ and\
		\citenamefont {Ma{\l}kiewicz}}]{BergeronMixmaster}%
	\BibitemOpen
	\bibfield  {author} {\bibinfo {author} {\bibfnamefont {H.}~\bibnamefont
			{Bergeron}}, \bibinfo {author} {\bibfnamefont {E.}~\bibnamefont {Czuchry}},
		\bibinfo {author} {\bibfnamefont {J.~P.}\ \bibnamefont {Gazeau}}, \ and\
		\bibinfo {author} {\bibfnamefont {P.}~\bibnamefont {Ma{\l}kiewicz}},\
	}\href@noop {} {\bibfield  {journal} {\bibinfo  {journal} {Universe}\
		}\textbf {\bibinfo {volume} {6}},\ \bibinfo {pages} {7} (\bibinfo {year}
		{2019})}\BibitemShut {NoStop}%
	\bibitem [{\citenamefont {G{\'o}{\'{z}}d{\'{z}}}\ \emph
		{et~al.}(2019)\citenamefont {G{\'o}{\'{z}}d{\'{z}}}, \citenamefont
		{Piechocki},\ and\ \citenamefont {Plewa}}]{GozdzBKL}%
	\BibitemOpen
	\bibfield  {author} {\bibinfo {author} {\bibfnamefont {A.}~\bibnamefont
			{G{\'o}{\'{z}}d{\'{z}}}}, \bibinfo {author} {\bibfnamefont {W.}~\bibnamefont
			{Piechocki}}, \ and\ \bibinfo {author} {\bibfnamefont {G.}~\bibnamefont
			{Plewa}},\ }\href {\doibase 10.1140/epjc/s10052-019-6571-4} {\bibfield
		{journal} {\bibinfo  {journal} {Eur. Phys. J. C}\ }\textbf {\bibinfo {volume}
			{79}},\ \bibinfo {pages} {45} (\bibinfo {year} {2019})}\BibitemShut {NoStop}%
	\bibitem [{\citenamefont {G{\'o}{\'z}d{\'z}}\ and\ \citenamefont
		{Piechocki}(2020)}]{GozdzBKL2}%
	\BibitemOpen
	\bibfield  {author} {\bibinfo {author} {\bibfnamefont {A.}~\bibnamefont
			{G{\'o}{\'z}d{\'z}}}\ and\ \bibinfo {author} {\bibfnamefont {W.}~\bibnamefont
			{Piechocki}},\ }\href {\doibase 10.1140/epjc/s10052-020-7668-5} {\bibfield
		{journal} {\bibinfo  {journal} {Eur. Phys. J. C}\ }\textbf {\bibinfo {volume}
			{80}},\ \bibinfo {pages} {142} (\bibinfo {year} {2020})}\BibitemShut
	{NoStop}%
	\bibitem [{\citenamefont {Casadio}(1998)}]{CasadioOS}%
	\BibitemOpen
	\bibfield  {author} {\bibinfo {author} {\bibfnamefont {R.}~\bibnamefont
			{Casadio}},\ }\href {\doibase 10.1103/PhysRevD.58.064013} {\bibfield
		{journal} {\bibinfo  {journal} {Phys. Rev. D}\ }\textbf {\bibinfo {volume}
			{58}},\ \bibinfo {pages} {064013} (\bibinfo {year} {1998})}\BibitemShut
	{NoStop}%
	\bibitem [{\citenamefont {Kelly}\ \emph {et~al.}()\citenamefont {Kelly},
		\citenamefont {Santacruz},\ and\ \citenamefont
		{Wilson-Ewing}}]{WilsonQuantumOS}%
	\BibitemOpen
	\bibfield  {author} {\bibinfo {author} {\bibfnamefont {J.~G.}\ \bibnamefont
			{Kelly}}, \bibinfo {author} {\bibfnamefont {R.}~\bibnamefont {Santacruz}}, \
		and\ \bibinfo {author} {\bibfnamefont {E.}~\bibnamefont {Wilson-Ewing}},\
	}\href@noop {} {}\Eprint {http://arxiv.org/abs/2006.09325} {arXiv:2006.09325
		[gr-qc]} \BibitemShut {NoStop}%
	\bibitem [{\citenamefont {Bambi}\ \emph {et~al.}(2014)\citenamefont {Bambi},
		\citenamefont {Malafarina},\ and\ \citenamefont {Modesto}}]{BambiBounce}%
	\BibitemOpen
	\bibfield  {author} {\bibinfo {author} {\bibfnamefont {C.}~\bibnamefont
			{Bambi}}, \bibinfo {author} {\bibfnamefont {D.}~\bibnamefont {Malafarina}}, \
		and\ \bibinfo {author} {\bibfnamefont {L.}~\bibnamefont {Modesto}},\ }\href
	{\doibase 10.1140/epjc/s10052-014-2767-9} {\bibfield  {journal} {\bibinfo
			{journal} {Eur. Phys. J. C}\ }\textbf {\bibinfo {volume} {74}},\ \bibinfo
		{pages} {2767} (\bibinfo {year} {2014})}\BibitemShut {NoStop}%
	\bibitem [{\citenamefont {Bambi}\ \emph {et~al.}(2013)\citenamefont {Bambi},
		\citenamefont {Malafarina},\ and\ \citenamefont
		{Modesto}}]{MalafarinaBounce}%
	\BibitemOpen
	\bibfield  {author} {\bibinfo {author} {\bibfnamefont {C.}~\bibnamefont
			{Bambi}}, \bibinfo {author} {\bibfnamefont {D.}~\bibnamefont {Malafarina}}, \
		and\ \bibinfo {author} {\bibfnamefont {L.}~\bibnamefont {Modesto}},\ }\href
	{\doibase 10.1103/PhysRevD.88.044009} {\bibfield  {journal} {\bibinfo
			{journal} {Phys. Rev. D}\ }\textbf {\bibinfo {volume} {88}},\ \bibinfo
		{pages} {044009} (\bibinfo {year} {2013})}\BibitemShut {NoStop}%
	\bibitem [{\citenamefont {Ashtekar}\ \emph {et~al.}(2018)\citenamefont
		{Ashtekar}, \citenamefont {Olmedo},\ and\ \citenamefont
		{Singh}}]{AshtekarBounce}%
	\BibitemOpen
	\bibfield  {author} {\bibinfo {author} {\bibfnamefont {A.}~\bibnamefont
			{Ashtekar}}, \bibinfo {author} {\bibfnamefont {J.}~\bibnamefont {Olmedo}}, \
		and\ \bibinfo {author} {\bibfnamefont {P.}~\bibnamefont {Singh}},\ }\href
	{\doibase 10.1103/PhysRevD.98.126003} {\bibfield  {journal} {\bibinfo
			{journal} {Phys. Rev. D}\ }\textbf {\bibinfo {volume} {98}},\ \bibinfo
		{pages} {126003} (\bibinfo {year} {2018})}\BibitemShut {NoStop}%
	\bibitem [{\citenamefont {Corichi}\ and\ \citenamefont
		{Singh}(2016)}]{KantowskiSachsBounce}%
	\BibitemOpen
	\bibfield  {author} {\bibinfo {author} {\bibfnamefont {A.}~\bibnamefont
			{Corichi}}\ and\ \bibinfo {author} {\bibfnamefont {P.}~\bibnamefont
			{Singh}},\ }\href {http://stacks.iop.org/0264-9381/33/i=5/a=055006}
	{\bibfield  {journal} {\bibinfo  {journal} {Class. Quantum Grav.}\ }\textbf
		{\bibinfo {volume} {33}},\ \bibinfo {pages} {055006} (\bibinfo {year}
		{2016})}\BibitemShut {NoStop}%
	\bibitem [{\citenamefont {Bojowald}(2020{\natexlab{a}})}]{BojowaldCritique}%
	\BibitemOpen
	\bibfield  {author} {\bibinfo {author} {\bibfnamefont {M.}~\bibnamefont
			{Bojowald}},\ }\href {\doibase 10.3390/universe6030036} {\bibfield  {journal}
		{\bibinfo  {journal} {Universe}\ }\textbf {\bibinfo {volume} {6}},\ \bibinfo
		{pages} {36} (\bibinfo {year} {2020}{\natexlab{a}})}\BibitemShut {NoStop}%
	\bibitem [{\citenamefont {Bojowald}(2020{\natexlab{b}})}]{BojowaldBHCritique}%
	\BibitemOpen
	\bibfield  {author} {\bibinfo {author} {\bibfnamefont {M.}~\bibnamefont
			{Bojowald}},\ }\href {\doibase 10.3390/universe6080125} {\bibfield  {journal}
		{\bibinfo  {journal} {Universe}\ }\textbf {\bibinfo {volume} {6}},\ \bibinfo
		{pages} {125} (\bibinfo {year} {2020}{\natexlab{b}})}\BibitemShut {NoStop}%
	\bibitem [{\citenamefont {Ambrus}\ and\ \citenamefont
		{H{\'a}j{\'i}\v{c}ek}(2005)}]{AmbrusHajicekLifetime}%
	\BibitemOpen
	\bibfield  {author} {\bibinfo {author} {\bibfnamefont {M.}~\bibnamefont
			{Ambrus}}\ and\ \bibinfo {author} {\bibfnamefont {P.}~\bibnamefont
			{H{\'a}j{\'i}\v{c}ek}},\ }\href {\doibase 10.1103/PhysRevD.72.064025}
	{\bibfield  {journal} {\bibinfo  {journal} {Phys. Rev. D}\ }\textbf {\bibinfo
			{volume} {72}},\ \bibinfo {pages} {064025} (\bibinfo {year}
		{2005})}\BibitemShut {NoStop}%
	\bibitem [{\citenamefont {Christodoulou}\ and\ \citenamefont
		{D'Ambrosio}()}]{ChristodoulouLifetime}%
	\BibitemOpen
	\bibfield  {author} {\bibinfo {author} {\bibfnamefont {M.}~\bibnamefont
			{Christodoulou}}\ and\ \bibinfo {author} {\bibfnamefont {F.}~\bibnamefont
			{D'Ambrosio}},\ }\href@noop {} {}\Eprint {http://arxiv.org/abs/1801.03027}
	{arXiv:1801.03027 [gr-qc]} \BibitemShut {NoStop}%
	%\%CITATION = ARXIV:1801.03027;\%\%
	\bibitem [{\citenamefont {Christodoulou}\ \emph {et~al.}(2016)\citenamefont
		{Christodoulou}, \citenamefont {Rovelli}, \citenamefont {Speziale},\ and\
		\citenamefont {Vilensky}}]{ChristodoulouLifetime2}%
	\BibitemOpen
	\bibfield  {author} {\bibinfo {author} {\bibfnamefont {M.}~\bibnamefont
			{Christodoulou}}, \bibinfo {author} {\bibfnamefont {C.}~\bibnamefont
			{Rovelli}}, \bibinfo {author} {\bibfnamefont {S.}~\bibnamefont {Speziale}}, \
		and\ \bibinfo {author} {\bibfnamefont {I.}~\bibnamefont {Vilensky}},\ }\href
	{\doibase 10.1103/PhysRevD.94.084035} {\bibfield  {journal} {\bibinfo
			{journal} {Phys. Rev. D}\ }\textbf {\bibinfo {volume} {94}},\ \bibinfo
		{pages} {084035} (\bibinfo {year} {2016})}\BibitemShut {NoStop}%
	\bibitem [{\citenamefont {Barcel{\'o}}\ \emph {et~al.}(2017)\citenamefont
		{Barcel{\'o}}, \citenamefont {Carballo-Rubio},\ and\ \citenamefont
		{Garay}}]{BarceloLifetime}%
	\BibitemOpen
	\bibfield  {author} {\bibinfo {author} {\bibfnamefont {C.}~\bibnamefont
			{Barcel{\'o}}}, \bibinfo {author} {\bibfnamefont {R.}~\bibnamefont
			{Carballo-Rubio}}, \ and\ \bibinfo {author} {\bibfnamefont {L.~J.}\
			\bibnamefont {Garay}},\ }\href
	{http://stacks.iop.org/0264-9381/34/i=10/a=105007} {\bibfield  {journal}
		{\bibinfo  {journal} {Class. Quantum Grav.}\ }\textbf {\bibinfo {volume}
			{34}},\ \bibinfo {pages} {105007} (\bibinfo {year} {2017})}\BibitemShut
	{NoStop}%
	\bibitem [{\citenamefont {Barcel{\'o}}\ \emph {et~al.}(2016)\citenamefont
		{Barcel{\'o}}, \citenamefont {Carballo-Rubio},\ and\ \citenamefont
		{Garay}}]{BarceloBounce1}%
	\BibitemOpen
	\bibfield  {author} {\bibinfo {author} {\bibfnamefont {C.}~\bibnamefont
			{Barcel{\'o}}}, \bibinfo {author} {\bibfnamefont {R.}~\bibnamefont
			{Carballo-Rubio}}, \ and\ \bibinfo {author} {\bibfnamefont {L.~J.}\
			\bibnamefont {Garay}},\ }\href {\doibase 10.1007/JHEP01(2016)157} {\bibfield
		{journal} {\bibinfo  {journal} {J. High Energy Phys.}\ }\textbf {\bibinfo
			{volume} {01}},\ \bibinfo {pages} {157} (\bibinfo {year} {2016})}\BibitemShut
	{NoStop}%
	\bibitem [{\citenamefont {Barcel{\'o}}\ \emph {et~al.}(2015)\citenamefont
		{Barcel{\'o}}, \citenamefont {Carballo-Rubio}, \citenamefont {Garay},\ and\
		\citenamefont {Jannes}}]{BarceloBounce2}%
	\BibitemOpen
	\bibfield  {author} {\bibinfo {author} {\bibfnamefont {C.}~\bibnamefont
			{Barcel{\'o}}}, \bibinfo {author} {\bibfnamefont {R.}~\bibnamefont
			{Carballo-Rubio}}, \bibinfo {author} {\bibfnamefont {L.~J.}\ \bibnamefont
			{Garay}}, \ and\ \bibinfo {author} {\bibfnamefont {G.}~\bibnamefont
			{Jannes}},\ }\href {http://stacks.iop.org/0264-9381/32/i=3/a=035012}
	{\bibfield  {journal} {\bibinfo  {journal} {Class. Quantum Grav.}\ }\textbf
		{\bibinfo {volume} {32}},\ \bibinfo {pages} {035012} (\bibinfo {year}
		{2015})}\BibitemShut {NoStop}%
	\bibitem [{\citenamefont {Barcel{\'o}}\ \emph {et~al.}(2014)\citenamefont
		{Barcel{\'o}}, \citenamefont {Carballo-Rubio},\ and\ \citenamefont
		{Garay}}]{BarceloBounce3}%
	\BibitemOpen
	\bibfield  {author} {\bibinfo {author} {\bibfnamefont {C.}~\bibnamefont
			{Barcel{\'o}}}, \bibinfo {author} {\bibfnamefont {R.}~\bibnamefont
			{Carballo-Rubio}}, \ and\ \bibinfo {author} {\bibfnamefont {L.~J.}\
			\bibnamefont {Garay}},\ }\href {\doibase 10.1142/S021827181442022X}
	{\bibfield  {journal} {\bibinfo  {journal} {Int. J. Mod. Phys. D}\ }\textbf
		{\bibinfo {volume} {23}},\ \bibinfo {pages} {1442022} (\bibinfo {year}
		{2014})}\BibitemShut {NoStop}%
	\bibitem [{\citenamefont {Liu}\ \emph {et~al.}(2014)\citenamefont {Liu},
		\citenamefont {Malafarina}, \citenamefont {Modesto},\ and\ \citenamefont
		{Bambi}}]{LiuBounce}%
	\BibitemOpen
	\bibfield  {author} {\bibinfo {author} {\bibfnamefont {Y.}~\bibnamefont
			{Liu}}, \bibinfo {author} {\bibfnamefont {D.}~\bibnamefont {Malafarina}},
		\bibinfo {author} {\bibfnamefont {L.}~\bibnamefont {Modesto}}, \ and\
		\bibinfo {author} {\bibfnamefont {C.}~\bibnamefont {Bambi}},\ }\href
	{\doibase 10.1103/PhysRevD.90.044040} {\bibfield  {journal} {\bibinfo
			{journal} {Phys. Rev. D}\ }\textbf {\bibinfo {volume} {90}},\ \bibinfo
		{pages} {044040} (\bibinfo {year} {2014})}\BibitemShut {NoStop}%
	\bibitem [{\citenamefont {Haggard}\ and\ \citenamefont
		{Rovelli}(2015)}]{HaggardRovelliBounce}%
	\BibitemOpen
	\bibfield  {author} {\bibinfo {author} {\bibfnamefont {H.~M.}\ \bibnamefont
			{Haggard}}\ and\ \bibinfo {author} {\bibfnamefont {C.}~\bibnamefont
			{Rovelli}},\ }\href {\doibase 10.1103/PhysRevD.92.104020} {\bibfield
		{journal} {\bibinfo  {journal} {Phys. Rev. D}\ }\textbf {\bibinfo {volume}
			{92}},\ \bibinfo {pages} {104020} (\bibinfo {year} {2015})}\BibitemShut
	{NoStop}%
	\bibitem [{\citenamefont {Malafarina}(2017)}]{MalafarinaBounceRev}%
	\BibitemOpen
	\bibfield  {author} {\bibinfo {author} {\bibfnamefont {D.}~\bibnamefont
			{Malafarina}},\ }\href {\doibase 10.3390/universe3020048} {\bibfield
		{journal} {\bibinfo  {journal} {Universe}\ }\textbf {\bibinfo {volume} {3}},\
		\bibinfo {pages} {48} (\bibinfo {year} {2017})}\BibitemShut {NoStop}%
	%\%CITATION = ARXIV:1703.04138;\%\%
	\bibitem [{\citenamefont {Bojowald}\ \emph {et~al.}(2008)\citenamefont
		{Bojowald}, \citenamefont {Harada},\ and\ \citenamefont
		{Tibrewala}}]{BojowaldLoopLTB}%
	\BibitemOpen
	\bibfield  {author} {\bibinfo {author} {\bibfnamefont {M.}~\bibnamefont
			{Bojowald}}, \bibinfo {author} {\bibfnamefont {T.}~\bibnamefont {Harada}}, \
		and\ \bibinfo {author} {\bibfnamefont {R.}~\bibnamefont {Tibrewala}},\ }\href
	{\doibase 10.1103/PhysRevD.78.064057} {\bibfield  {journal} {\bibinfo
			{journal} {Phys. Rev. D}\ }\textbf {\bibinfo {volume} {78}},\ \bibinfo
		{pages} {064057} (\bibinfo {year} {2008})}\BibitemShut {NoStop}%
	\bibitem [{\citenamefont {Bojowald}\ \emph {et~al.}(2009)\citenamefont
		{Bojowald}, \citenamefont {Reyes},\ and\ \citenamefont
		{Tibrewala}}]{BojowaldLoopLTB2}%
	\BibitemOpen
	\bibfield  {author} {\bibinfo {author} {\bibfnamefont {M.}~\bibnamefont
			{Bojowald}}, \bibinfo {author} {\bibfnamefont {J.~D.}\ \bibnamefont {Reyes}},
		\ and\ \bibinfo {author} {\bibfnamefont {R.}~\bibnamefont {Tibrewala}},\
	}\href {\doibase 10.1103/PhysRevD.80.084002} {\bibfield  {journal} {\bibinfo
			{journal} {Phys. Rev. D}\ }\textbf {\bibinfo {volume} {80}},\ \bibinfo
		{pages} {084002} (\bibinfo {year} {2009})}\BibitemShut {NoStop}%
	\bibitem [{\citenamefont {De~Lorenzo}\ and\ \citenamefont
		{Perez}(2016)}]{DeLorenzoFireworks}%
	\BibitemOpen
	\bibfield  {author} {\bibinfo {author} {\bibfnamefont {T.}~\bibnamefont
			{De~Lorenzo}}\ and\ \bibinfo {author} {\bibfnamefont {A.}~\bibnamefont
			{Perez}},\ }\href {\doibase 10.1103/PhysRevD.93.124018} {\bibfield  {journal}
		{\bibinfo  {journal} {Phys. Rev. D}\ }\textbf {\bibinfo {volume} {93}},\
		\bibinfo {pages} {124018} (\bibinfo {year} {2016})}\BibitemShut {NoStop}%
	\bibitem [{\citenamefont {Rovelli}\ and\ \citenamefont
		{Martin-Dussaud}(2018)}]{RovelliFireworks}%
	\BibitemOpen
	\bibfield  {author} {\bibinfo {author} {\bibfnamefont {C.}~\bibnamefont
			{Rovelli}}\ and\ \bibinfo {author} {\bibfnamefont {P.}~\bibnamefont
			{Martin-Dussaud}},\ }\href {\doibase 10.1088/1361-6382/aacb74} {\bibfield
		{journal} {\bibinfo  {journal} {Classical and Quantum Gravity}\ }\textbf
		{\bibinfo {volume} {35}},\ \bibinfo {pages} {147002} (\bibinfo {year}
		{2018})}\BibitemShut {NoStop}%
	\bibitem [{\citenamefont {Martin-Dussaud}\ and\ \citenamefont
		{Rovelli}(2019)}]{MartinDussaudFireworks}%
	\BibitemOpen
	\bibfield  {author} {\bibinfo {author} {\bibfnamefont {P.}~\bibnamefont
			{Martin-Dussaud}}\ and\ \bibinfo {author} {\bibfnamefont {C.}~\bibnamefont
			{Rovelli}},\ }\href {\doibase 10.1088/1361-6382/ab5097} {\bibfield  {journal}
		{\bibinfo  {journal} {Classical and Quantum Gravity}\ }\textbf {\bibinfo
			{volume} {36}},\ \bibinfo {pages} {245002} (\bibinfo {year}
		{2019})}\BibitemShut {NoStop}%
	\bibitem [{\citenamefont {{Ben Achour}}\ \emph {et~al.}(2020)\citenamefont
		{{Ben Achour}}, \citenamefont {Brahma},\ and\ \citenamefont
		{Uzan}}]{AchourBounce1}%
	\BibitemOpen
	\bibfield  {author} {\bibinfo {author} {\bibfnamefont {J.}~\bibnamefont {{Ben
					Achour}}}, \bibinfo {author} {\bibfnamefont {S.}~\bibnamefont {Brahma}}, \
		and\ \bibinfo {author} {\bibfnamefont {J.-P.}\ \bibnamefont {Uzan}},\ }\href
	{\doibase 10.1088/1475-7516/2020/03/041} {\bibfield  {journal} {\bibinfo
			{journal} {J. Cosmol. and Astropart Phys.}\ }\textbf {\bibinfo {volume}
			{2020}},\ \bibinfo {pages} {041} (\bibinfo {year} {2020})}\BibitemShut
	{NoStop}%
	\bibitem [{\citenamefont {{Ben Achour}}\ and\ \citenamefont
		{Uzan}()}]{AchourBounce2}%
	\BibitemOpen
	\bibfield  {author} {\bibinfo {author} {\bibfnamefont {J.}~\bibnamefont {{Ben
					Achour}}}\ and\ \bibinfo {author} {\bibfnamefont {J.-P.}\ \bibnamefont
			{Uzan}},\ }\href@noop {} {}\Eprint {http://arxiv.org/abs/2001.06153}
	{arXiv:2001.06153 [gr-qc]} \BibitemShut {NoStop}%
	\bibitem [{\citenamefont {{Ben Achour}}\ \emph {et~al.}()\citenamefont {{Ben
				Achour}}, \citenamefont {Brahma}, \citenamefont {Mukohyama},\ and\
		\citenamefont {Uzan}}]{AchourBounce3}%
	\BibitemOpen
	\bibfield  {author} {\bibinfo {author} {\bibfnamefont {J.}~\bibnamefont {{Ben
					Achour}}}, \bibinfo {author} {\bibfnamefont {S.}~\bibnamefont {Brahma}},
		\bibinfo {author} {\bibfnamefont {S.}~\bibnamefont {Mukohyama}}, \ and\
		\bibinfo {author} {\bibfnamefont {J.-P.}\ \bibnamefont {Uzan}},\ }\href@noop
	{} {}\Eprint {http://arxiv.org/abs/2004.12977} {arXiv:2004.12977 [gr-qc]}
	\BibitemShut {NoStop}%
	\bibitem [{\citenamefont {Münch}()}]{MuenchMatching}%
	\BibitemOpen
	\bibfield  {author} {\bibinfo {author} {\bibfnamefont {J.}~\bibnamefont
			{Münch}},\ }\href@noop {} {}\Eprint {http://arxiv.org/abs/2010.13480}
	{arXiv:2010.13480 [gr-qc]} \BibitemShut {NoStop}%
	\bibitem [{\citenamefont {Hayward}(2006)}]{HaywardRegular}%
	\BibitemOpen
	\bibfield  {author} {\bibinfo {author} {\bibfnamefont {S.~A.}\ \bibnamefont
			{Hayward}},\ }\href {\doibase 10.1103/PhysRevLett.96.031103} {\bibfield
		{journal} {\bibinfo  {journal} {Phys. Rev. Lett.}\ }\textbf {\bibinfo
			{volume} {96}},\ \bibinfo {pages} {031103} (\bibinfo {year}
		{2006})}\BibitemShut {NoStop}%
	\bibitem [{\citenamefont {Carballo-Rubio}\ \emph {et~al.}(2020)\citenamefont
		{Carballo-Rubio}, \citenamefont {Filippo}, \citenamefont {Liberati},\ and\
		\citenamefont {Visser}}]{CarballoRubioRegular}%
	\BibitemOpen
	\bibfield  {author} {\bibinfo {author} {\bibfnamefont {R.}~\bibnamefont
			{Carballo-Rubio}}, \bibinfo {author} {\bibfnamefont {F.~D.}\ \bibnamefont
			{Filippo}}, \bibinfo {author} {\bibfnamefont {S.}~\bibnamefont {Liberati}}, \
		and\ \bibinfo {author} {\bibfnamefont {M.}~\bibnamefont {Visser}},\ }\href
	{\doibase 10.1088/1361-6382/ab8141} {\bibfield  {journal} {\bibinfo
			{journal} {Classical and Quantum Gravity}\ }\textbf {\bibinfo {volume}
			{37}},\ \bibinfo {pages} {145005} (\bibinfo {year} {2020})}\BibitemShut
	{NoStop}%
	\bibitem [{\citenamefont {Ansoldi}()}]{AnsoldiRegularRev}%
	\BibitemOpen
	\bibfield  {author} {\bibinfo {author} {\bibfnamefont {S.}~\bibnamefont
			{Ansoldi}},\ }\href@noop {} {}\Eprint {http://arxiv.org/abs/0802.0330}
	{arXiv:0802.0330 [gr-qc]} \BibitemShut {NoStop}%
	\bibitem [{\citenamefont {Martel}\ and\ \citenamefont
		{Poisson}(2001)}]{MartelCoordinates}%
	\BibitemOpen
	\bibfield  {author} {\bibinfo {author} {\bibfnamefont {K.}~\bibnamefont
			{Martel}}\ and\ \bibinfo {author} {\bibfnamefont {E.}~\bibnamefont
			{Poisson}},\ }\href {\doibase 10.1119/1.1336836} {\bibfield  {journal}
		{\bibinfo  {journal} {Am. J. Phys.}\ }\textbf {\bibinfo {volume} {69}},\
		\bibinfo {pages} {476} (\bibinfo {year} {2001})}\BibitemShut {NoStop}%
	\bibitem [{\citenamefont {Gautreau}\ and\ \citenamefont
		{Hoffmann}(1978)}]{GautreauCoordinates}%
	\BibitemOpen
	\bibfield  {author} {\bibinfo {author} {\bibfnamefont {R.}~\bibnamefont
			{Gautreau}}\ and\ \bibinfo {author} {\bibfnamefont {B.}~\bibnamefont
			{Hoffmann}},\ }\href {\doibase 10.1103/PhysRevD.17.2552} {\bibfield
		{journal} {\bibinfo  {journal} {Phys. Rev. D}\ }\textbf {\bibinfo {volume}
			{17}},\ \bibinfo {pages} {2552} (\bibinfo {year} {1978})}\BibitemShut
	{NoStop}%
	\bibitem [{\citenamefont {Faraoni}(2015)}]{FaraoniHorizons}%
	\BibitemOpen
	\bibfield  {author} {\bibinfo {author} {\bibfnamefont {V.}~\bibnamefont
			{Faraoni}},\ }\href {https://books.google.de/books?id=iK8YCgAAQBAJ} {\emph
		{\bibinfo {title} {Cosmological and Black Hole Apparent Horizons}}},\ Lecture
	Notes in Physics\ (\bibinfo  {publisher} {Springer International
		Publishing},\ \bibinfo {year} {2015})\BibitemShut {NoStop}%
	\bibitem [{\citenamefont {Faraoni}\ and\ \citenamefont
		{Giusti}(2020)}]{FaraoniUnphysical1}%
	\BibitemOpen
	\bibfield  {author} {\bibinfo {author} {\bibfnamefont {V.}~\bibnamefont
			{Faraoni}}\ and\ \bibinfo {author} {\bibfnamefont {A.}~\bibnamefont
			{Giusti}},\ }\href {\doibase 10.3390/sym12081264} {\bibfield  {journal}
		{\bibinfo  {journal} {Symmetry}\ }\textbf {\bibinfo {volume} {12}},\ \bibinfo
		{pages} {1264} (\bibinfo {year} {2020})}\BibitemShut {NoStop}%
	\bibitem [{\citenamefont {Faraoni}\ \emph {et~al.}()\citenamefont {Faraoni},
		\citenamefont {Giusti},\ and\ \citenamefont {Bean}}]{FaraoniUnphysical2}%
	\BibitemOpen
	\bibfield  {author} {\bibinfo {author} {\bibfnamefont {V.}~\bibnamefont
			{Faraoni}}, \bibinfo {author} {\bibfnamefont {A.}~\bibnamefont {Giusti}}, \
		and\ \bibinfo {author} {\bibfnamefont {T.~F.}\ \bibnamefont {Bean}},\
	}\href@noop {} {}\Eprint {http://arxiv.org/abs/2010.00069} {arXiv:2010.00069
		[gr-qc]} \BibitemShut {NoStop}%
	\bibitem [{\citenamefont {Schmitz}(2020)}]{MeOS}%
	\BibitemOpen
	\bibfield  {author} {\bibinfo {author} {\bibfnamefont {T.}~\bibnamefont
			{Schmitz}},\ }\href {\doibase 10.1103/PhysRevD.101.026016} {\bibfield
		{journal} {\bibinfo  {journal} {Phys. Rev. D}\ }\textbf {\bibinfo {volume}
			{101}},\ \bibinfo {pages} {026016} (\bibinfo {year} {2020})}\BibitemShut
	{NoStop}%
	\bibitem [{\citenamefont {Schindler}\ and\ \citenamefont
		{Aguirre}(2018)}]{SchindlerDiagram}%
	\BibitemOpen
	\bibfield  {author} {\bibinfo {author} {\bibfnamefont {J.~C.}\ \bibnamefont
			{Schindler}}\ and\ \bibinfo {author} {\bibfnamefont {A.}~\bibnamefont
			{Aguirre}},\ }\href {\doibase 10.1088/1361-6382/aabce2} {\bibfield  {journal}
		{\bibinfo  {journal} {Class. Quant. Grav.}\ }\textbf {\bibinfo {volume}
			{35}},\ \bibinfo {pages} {105019} (\bibinfo {year} {2018})}\BibitemShut
	{NoStop}%
	\bibitem [{\citenamefont {Eardley}(1974)}]{EardleyInstability}%
	\BibitemOpen
	\bibfield  {author} {\bibinfo {author} {\bibfnamefont {D.~M.}\ \bibnamefont
			{Eardley}},\ }\href {\doibase 10.1103/PhysRevLett.33.442} {\bibfield
		{journal} {\bibinfo  {journal} {Phys. Rev. Lett.}\ }\textbf {\bibinfo
			{volume} {33}},\ \bibinfo {pages} {442} (\bibinfo {year} {1974})}\BibitemShut
	{NoStop}%
	\bibitem [{\citenamefont {Blau}(1989)}]{BlauInstability}%
	\BibitemOpen
	\bibfield  {author} {\bibinfo {author} {\bibfnamefont {S.~K.}\ \bibnamefont
			{Blau}},\ }\href {\doibase 10.1103/PhysRevD.39.2901} {\bibfield  {journal}
		{\bibinfo  {journal} {Phys. Rev. D}\ }\textbf {\bibinfo {volume} {39}},\
		\bibinfo {pages} {2901} (\bibinfo {year} {1989})}\BibitemShut {NoStop}%
	\bibitem [{\citenamefont {Barrab\`es}\ \emph {et~al.}(1993)\citenamefont
		{Barrab\`es}, \citenamefont {Brady},\ and\ \citenamefont
		{Poisson}}]{BarrabesInstability}%
	\BibitemOpen
	\bibfield  {author} {\bibinfo {author} {\bibfnamefont {C.}~\bibnamefont
			{Barrab\`es}}, \bibinfo {author} {\bibfnamefont {P.~R.}\ \bibnamefont
			{Brady}}, \ and\ \bibinfo {author} {\bibfnamefont {E.}~\bibnamefont
			{Poisson}},\ }\href {\doibase 10.1103/PhysRevD.47.2383} {\bibfield  {journal}
		{\bibinfo  {journal} {Phys. Rev. D}\ }\textbf {\bibinfo {volume} {47}},\
		\bibinfo {pages} {2383} (\bibinfo {year} {1993})}\BibitemShut {NoStop}%
	\bibitem [{\citenamefont {Ori}\ and\ \citenamefont
		{Poisson}(1994)}]{OriInstability}%
	\BibitemOpen
	\bibfield  {author} {\bibinfo {author} {\bibfnamefont {A.}~\bibnamefont
			{Ori}}\ and\ \bibinfo {author} {\bibfnamefont {E.}~\bibnamefont {Poisson}},\
	}\href {\doibase 10.1103/PhysRevD.50.6150} {\bibfield  {journal} {\bibinfo
			{journal} {Phys. Rev. D}\ }\textbf {\bibinfo {volume} {50}},\ \bibinfo
		{pages} {6150} (\bibinfo {year} {1994})}\BibitemShut {NoStop}%
	\bibitem [{\citenamefont {Lake}\ and\ \citenamefont
		{Roeder}(1976)}]{LakeInstability}%
	\BibitemOpen
	\bibfield  {author} {\bibinfo {author} {\bibfnamefont {K.}~\bibnamefont
			{Lake}}\ and\ \bibinfo {author} {\bibfnamefont {R.~C.}\ \bibnamefont
			{Roeder}},\ }\href {\doibase 10.1007/BF02719664} {\bibfield  {journal}
		{\bibinfo  {journal} {Lett. Nuovo Cimento}\ }\textbf {\bibinfo {volume}
			{16}},\ \bibinfo {pages} {17} (\bibinfo {year} {1976})}\BibitemShut {NoStop}%
	\bibitem [{\citenamefont {Lake}(1978)}]{LakeInstability2}%
	\BibitemOpen
	\bibfield  {author} {\bibinfo {author} {\bibfnamefont {K.}~\bibnamefont
			{Lake}},\ }\href {\doibase 10.1038/272599a0} {\bibfield  {journal} {\bibinfo
			{journal} {Nature}\ }\textbf {\bibinfo {volume} {272}},\ \bibinfo {pages}
		{599} (\bibinfo {year} {1978})}\BibitemShut {NoStop}%
	\bibitem [{\citenamefont {Poisson}(2004)}]{PoissonRelativistsToolkit}%
	\BibitemOpen
	\bibfield  {author} {\bibinfo {author} {\bibfnamefont {E.}~\bibnamefont
			{Poisson}},\ }\href {https://books.google.de/books?id=bk2XEgz\_ML4C} {\emph
		{\bibinfo {title} {A Relativist's Toolkit: The Mathematics of Black-Hole
				Mechanics}}}\ (\bibinfo  {publisher} {Cambridge University Press,
		Cambridge},\ \bibinfo {year} {2004})\BibitemShut {NoStop}%
\end{thebibliography}

%copy bibliography in here
%merlin.mbs apsrev4-1.bst 2010-07-25 4.21a (PWD, AO, DPC) hacked
%Control: key (0)
%Control: author (8) initials jnrlst
%Control: editor formatted (1) identically to author
%Control: production of article title (-1) disabled
%Control: page (0) single
%Control: year (1) truncated
%Control: production of eprint (0) enabled
%

%%%%%%%%%%%%%%%%%%%%%%%%%%%%%%%%%%%%%%%%%%%%%%%%%%%%%%%%%%%%%%%%%

\end{document}